\documentclass[sigplan, twocolumn, nonacm]{acmart}

\AtBeginDocument{%
  }

\setcopyright{none} 

\usepackage{amsmath}
\usepackage{amsfonts} 
\usepackage{algorithm}
\usepackage{algpseudocode}
\usepackage{graphicx}
\usepackage{textcomp}
\usepackage{xcolor}
\usepackage[keeplastbox]{flushend}
\usepackage{fancyhdr}
\usepackage{booktabs} 
\usepackage{multirow} 
\usepackage{soul} 
\usepackage{xspace}
\usepackage{balance}
\usepackage{subfigure}
\usepackage{kotex} 
\usepackage{multicol} 

\newcommand{\abcmem}{\textsc{H2M2}\xspace}
\newcommand{\amalgam}{\textsc{H2M2}\xspace}

\begin{document}

\title{Hardware-based Heterogeneous Memory Management for Large Language Model Inference}

\author{Soojin Hwang}
\affiliation{%
    \institution{KAIST}
    \country{Republic of Korea}
}
\email{sjhwang@casys.kaist.ac.kr}

\author{Jungwoo Kim}
\affiliation{%
    \institution{Stanford University}
    \country{California, USA}
}
\email{jungwkim@stanford.edu}

\author{Sanghyeon Lee}
\affiliation{%
    \institution{KAIST}
    \country{Republic of Korea}
}
\email{leesh6796@casys.kaist.ac.kr}

\author{Hongbeen Kim}
\affiliation{%
    \institution{KAIST}
    \country{Republic of Korea}
}
\email{hbkim@casys.kaist.ac.kr}

\author{Jaehyuk Huh}
\affiliation{%
    \institution{KAIST}
    \country{Republic of Korea}
}
\email{jhhuh@kaist.ac.kr}

\begin{abstract}
A large language model (LLM) is one of the most important emerging machine learning applications nowadays. However, due to its huge model size and runtime increase of the memory footprint, LLM inferences suffer from the lack of memory capacity in conventional systems consisting of multiple GPUs with a modest amount of high bandwidth memory. Moreover, since LLM contains many bandwidth-intensive kernels, only focusing on the memory capacity without considering the bandwidth incurs a serious performance degradation. To handle such conflicting memory capacity and bandwidth demands in a cost-effective way, this study investigates the potential of heterogeneous memory systems, proposing \amalgam. It uses an asymmetric memory architecture consisting of capacity-centric and bandwidth-centric memory with computation units attached to each memory device. With the asymmetric memory, we first analyze the effect of kernel-memory mapping for the asymmetric memory. Second, we propose a dynamic runtime algorithm that finds a mapping solution considering the characteristics of LLM operations and the change of footprint during LLM inference. Third, we advocate the need for memory abstraction for the efficient management of the asymmetric memory. \amalgam outperforms the conventional homogeneous memory system with LPDDR by 1.46$\times$, 1.55$\times$, and 2.94$\times$ speedup in GPT3-175B, Chinchilla-70B, and Llama2-70B, respectively.
\end{abstract}

\keywords{Large Language Models (LLMs), Heterogeneous memory system, Hardware accelerators}

\maketitle

\pagenumbering{arabic}

\section{Introduction}

Recently, decoder-only transformer-based large language models (LLMs) are widely used for their simple structure and powerful performance~\cite{palm, bloom, gpt3, opt, llama2}.
However, the characteristics of LLMs make it challenging to accelerate their computation with conventional systems:
First, the large sizes of model parameters and activations require significant memory capacity and the infrequent tensor reuse leads to low locality, demanding high memory bandwidth.
Second, decoder-based LLMs store the KV (Key-Value) cache to prevent repetitive computation, which expands in size with each token generation. 
Unlike the prior ML inferences, this leads to significant dynamic changes in memory footprint during inference runtime.
Third, the opportunity for batching is limited, as the attention layer cannot be batched with multiple requests.
Furthermore, the attention layer of a token generation phase contains numerous memory intensive GEMV (GEneral Matrix-Vector multiplication) kernels.

To address the substantial memory footprint and high bandwidth requirement, multi-GPU systems are extensively used for LLM acceleration. 
However, the limited locality of memory accesses presents difficulties in fully utilizing the high computation power of GPUs, because multiple GPUs are needed to store weights and KV cache in relative small HBMs (High Bandwidth Memory). The capacity requirement of LLMs increases LLM service costs significantly with many GPUs necessary to store the weights and KV cache in a distributed way.  
Furthermore, applying the model/tensor parallelism to multi-GPU systems incurs considerable communication and synchronization overheads~\cite{deepspeed, petals}.

To alleviate the memory capacity challenge of LLMs, recent studies used the host memory as the second-level memory for the KV cache and/or weights~\cite{alisa, flexgen}. However, these approaches incur significant performance degradation since they need to access CPU-side memory through limited PCIe bandwidth.
A recent alternative HW-based solution integrates capacity-centric memory combined with a custom accelerator chip~\cite{cxlpnm}. 
The approach showed that an accelerator with a large capacity of LPDDR memory can outperform multi-GPU systems in LLM performance for non-batched operations, as using many GPUs for larger models increases communication and synchronization overheads~\cite{cxlpnm}.
Although this approach accommodates the capacity demand of LLM, the low bandwidth of LPDDR memory still limits the performance of LLM computation.

\begin{figure}[t]
    \centering
    \includegraphics[width=\columnwidth]{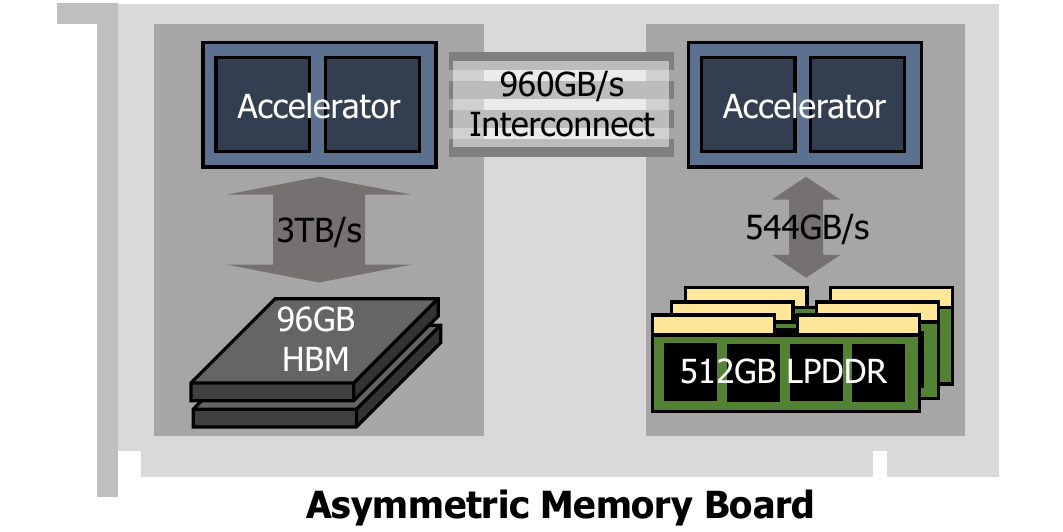}
    \vspace{-0.2in}
    \caption{Accelerator substrate used by \amalgam.}
    \vspace{-0.2in}
    \label{fig:overview}
\end{figure}

Two major memory demands of LLMs, \textit{large capacity} and \textit{high bandwidth} must be satisfied to improve the performance by overcoming the limitation of prior approaches.
To reach the goal, this paper investigates a heterogeneous memory architecture with the capacity-centric memory module and the bandwidth-centric memory module.
However, the strict hierarchical organization of conventional heterogeneous memory which uses the high bandwidth memory as the first level memory backed by the low bandwidth memory is not suitable for LLMs, as the low locality of LLMs makes the performance bounded by the low bandwidth of capacity-centric memory. 

In this paper, we employ an \textit{asymmetric memory} architecture for LLM acceleration combining the bandwidth-centric and capacity-centric memory modules in a parallel way, as supported in a recent Grace Hopper architecture~\cite{gracehopper}.
The asymmetric memory adds accelerators to both of the bandwidth-centric and capacity-centric memory modules.
This study investigates how the mapping of data and computation of LLMs can properly exploit the potential of the asymmetric memory architecture and what hardware supports are needed to hide the complexity of asymmetric memory. We explore mapping techniques such as kernel fusion and layer split to the asymmetric memory, and analyze patterns of the best \textit{Kernel-Memory Mapping} to unleash the potential of performance.

This paper proposes a dynamic mapping algorithm for accelerator-based asymmetric memory system with hardware support for memory abstraction, called \amalgam.
Based on the best mapping analysis, we propose a runtime policy for the kernel-memory mapping, which can be achieved by solving a simple linear problem. It can adjust the memory mapping effectively when the KV cache size dynamically changes for different batch sizes and token lengths.
In addition, we propose an efficient dynamic memory management for asymmetric memory to address the KV cache memory allocation problem raised by a recent study~\cite{vllm}, and to provide a unified memory view under dynamic changes of the memory mapping. 

We evaluate the performance of \abcmem using a cycle-accurate simulator for generation phases of three LLMs: GPT3-175B, Chinchilla-70B, and Llama2-70B.
\amalgam outperforms the capacity-centric memory architecture with the same number of computation units by 1.46$\times$, 1.55$\times$, and 2.94$\times$ speedup for GPT3-175B, Chinchilla-70B, and Llama2-70B, respectively.
The hardware support for memory abstraction of \amalgam incurs up to only 1.36\% performance overhead in three LLM workloads.
The kernel-memory mapping based on the greedy mapping policy of \abcmem shows less than 5\% additional degradation compared to the optimal kernel-memory mapping strategy in all three LLM workloads, which is negligible for the overall performance gain from asymmetric memory architecture.

The contribution of this paper are as follows:
\begin{itemize}
    \item This study applies an asymmetric memory architecture to LLM acceleration, and analyzes the effect of kernel-memory mapping policies to find the best one.
    \item It proposes a near-optimal runtime algorithm under dynamically changing KV cache sizes.
    \item It proposes an efficient memory abstraction scheme to address the challenges for the KV cache size and mapping changes.
\end{itemize}
\section{Background}
\subsection{Large Language Models}

Language modeling entails predicting the forthcoming sequence of output tokens (typically words or subwords) from given a set of input tokens.
Recently, transformer-based generative language models have been extensively explored owing to their inherent scalability and self-attention mechanisms~\cite{attention}.
These transformer-based language models with a significant number of parameters are commonly known as Large Language Models (LLMs)~\cite{llm-survey}.

\begin{figure}[t]
    \centering
    \includegraphics[width=\columnwidth]{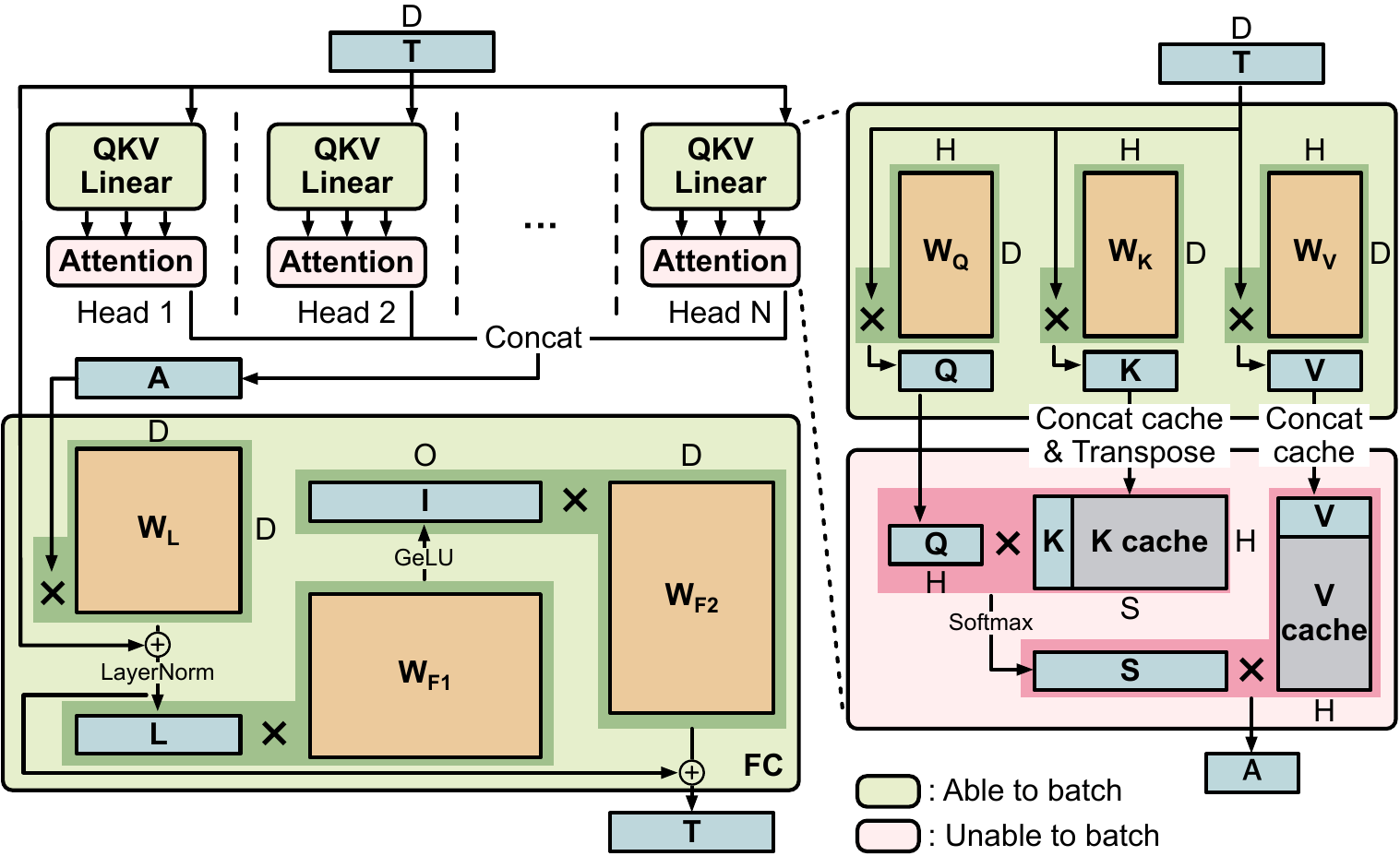}
    \vspace{-0.3in}
    \caption{The topology of the decoder layer in GPT3. Orange boxes with \texttt{W} labels represent the weight parameter tensors, blue boxes represent the input activation tensors, and gray boxes represent the KV cache tensors.}
    \vspace{-0.1in}
    \label{fig:decoder_topology}
\end{figure}

\noindent
{\bf Decoder-based LLM:}
Recent LLMs are based on on a \textit{transformer decoder layer} that generates subsequent tokens autoregressively (\textit{generation phase}), after processing the prompt with multiple tokens (\textit{prompt phase})~\cite{gpt3, gpt4, bloom, llama, llama2, palm, palm2}.
These LLMs are constructed with multiple decoder layers, each sharing the same topology in terms of tensor dimensions and operations, albeit with different parameter values.
%
%
Figure~\ref{fig:decoder_topology} illustrates the decoder layer topology for the autoregressive generation phase in the GPT3 model, which dominates inference execution time~\cite{gpt3}. Key hyperparameters include $H$, the dimension of each attention head, $N$, the number of heads, $D$, the dimension of the bottleneck layer following multi-head attention, and $O$, the dimension of the feed-forward layer. The sequence length $S$ represents the total length of the prompt and previously generated tokens.

As illustrated in Figure~\ref{fig:decoder_topology}, the decoder in the generation phase includes green-colored batching-compatible operations and pink-colored batching-incompatible operations. Batching-compatible operations are computed as GEneral Matrix-Matrix multiplication (GEMM) kernels when the batch size is greater than one, or as GEneral Matrix-Vector multiplication (GEMV) kernels for a batch size of one. In contrast, batching-incompatible operations are always computed as GEMV kernels, regardless of batch size. In the rest of this paper, we classify these decoder kernels into three groups: \textit{qkv-linear}, \textit{attention}, and \textit{fc} layers.
During the computation of transformer decoder, the embedding vector \texttt{T} is first passed to \textit{qkv-linear} sublayer to generate LLM contexts: Query (\texttt{Q}), key (\texttt{K}), and value (\texttt{V}).
Next, \textit{attention} sublayer computes attention value \texttt{A} using \texttt{Q}, \texttt{K}, and \texttt{V}. This sublayer utilizes not only the contexts of the current input token, but also the contexts of prior tokens - called as KV cache.
Lastly, \textit{fc} sublayer computes the final output, and this sublayer includes projection and feed-forward network.
While the weight - activation multiplication kernels of \textit{qkv-linear} and \textit{fc} can be batched with multiple requests, it is usually infeasible to batch KV cache - activation multiplication kernels in \textit{attention}.
This is because of the nature of the KV cache; Due to the difference of user requirements between requests, requests in a batch rarely have common values in the KV cache, not allowing the batched computation.

\subsection{Challenges of Accelerating LLM Inference}


\begin{figure}[t]
    \centering
    \includegraphics[width=\columnwidth]{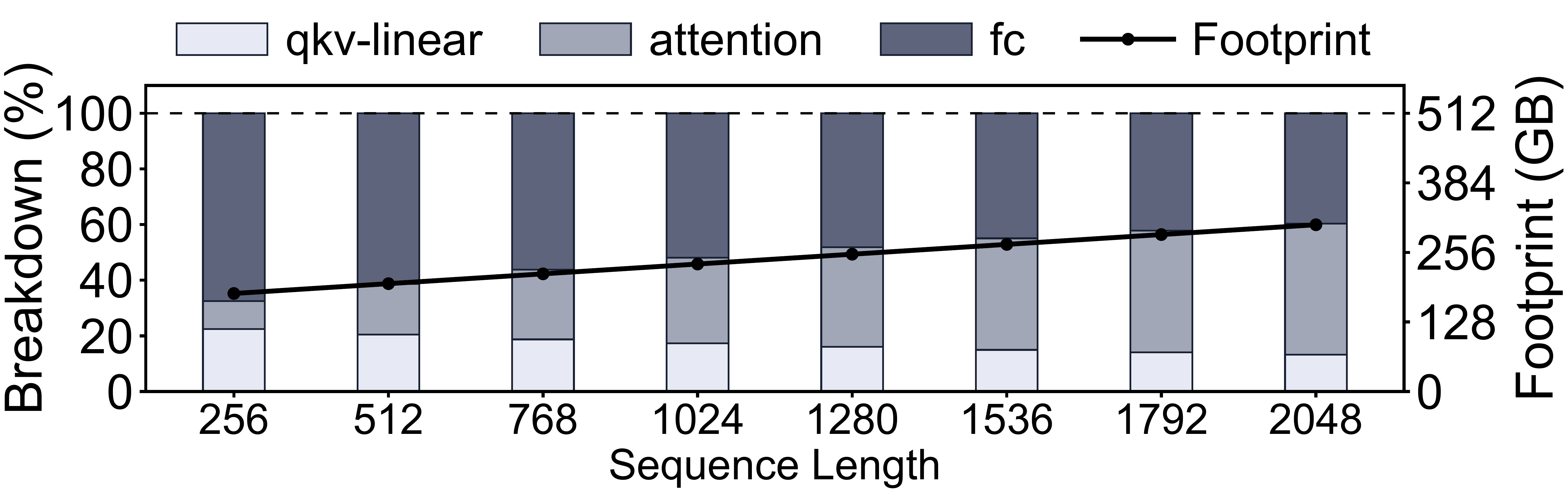}
    \vspace{-4ex}
    \caption{Footprint breakdown for batch size 32, sequence length growing from 256 to 2048 in GPT3-175B. The portion of \textit{attention} increases due to the increase of KV cache size.}
    \label{fig:footprint-breakdown}
    \vspace{-2ex}
\end{figure}

\subsubsection{Large Footprint and Limited Locality}
\label{subsubsec:footprint}


\begin{figure*}[t]
    \centering
    \includegraphics[width=\textwidth]{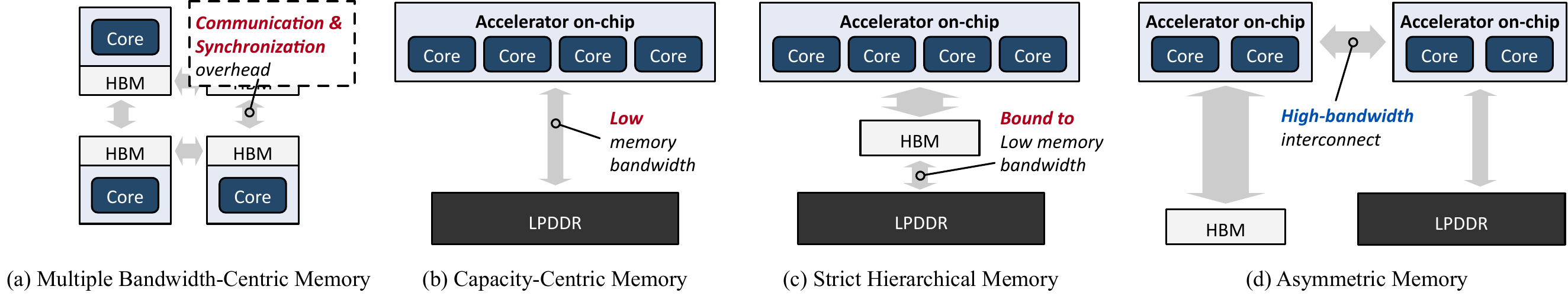}
    \vspace{-5ex}
    \caption{Four possible configurations of memory systems applicable for LLM inference acceleration.}
    \vspace{-2ex}
    \label{fig:hetero_system}
\end{figure*}

Modern LLMs often require hundreds of gigabytes~\cite{gpt3, palm, bloom, opt}.
Moreover, the rarity of data repetition among kernels and layers leads to strict dependencies among operations, as described in Figure~\ref{fig:decoder_topology}.
The considerable memory footprint and strict dependencies in LLMs contribute to excessively long data reuse distances, which limits temporal locality.
For instance, in GPT3-175B with FP16 precision, the reuse distance of weight parameters is at least 350GB.
Several prior works have introduced the methodology of increasing inter-kernel temporal locality through batching, but their limited data reuse still results in the lack of temporal locality for the overall model~\cite{orca, vllm}.

\subsubsection{Unique Characteristics of KV Cache}
As depicted in Figure~\ref{fig:decoder_topology}, the \textit{attention} key and value tensors are preserved in memory for future reuse, forming what is known as \textit{KV cache}~\cite{kvcache}.
KV cache continuously expands throughout the auto-regressive generation phase until the generation of the end-of-sequence (eos) token.
Therefore, as shown in the Figure~\ref{fig:footprint-breakdown}, the model footprint dynamically changes during inference, resulting in significant storage overhead.
Thus, accelerating LLM inference requires a unique approach distinct from traditional DNN models.
For example, vLLM introduced efficient memory management for key and value tensors within \textit{attention} layers by employing virtual memory and paging techniques with software supports~\cite{vllm}.
However, vLLM presupposed that GPU device memory is large enough to store all weight parameters, still facing the limitation when running large models.

\subsubsection{Limitation on Batching}

GEMV serves as a basic building block of the decoder due to batching-incompatible operations. However, its arithmetic intensity (i.e. \# of operations per memory traffic) is $O(1)$, whereas GEMM has an arithmetic intensity of $O(n)$.
Due to limited data reuse, GPUs handle GEMV less efficiently than GEMM, making it necessary to convert GEMV into GEMM through batching to fully utilize GPU computation units. To address this, Orca selectively batches compatible operators at the iteration level~\cite{orca}.
However, batching requests within \textit{attention} layers is nearly impossible, as these layers do not have weight parameters and compute each request independently using inputs such as query, key, and value. Prior studies have proposed integrating GEMV-optimized processing-in-memory (PIM) technology to efficiently accelerate attention layers~\cite{neupims, ianus, attacc}. However, PIM architectures are limited by their small memory capacity due to the inclusion of computation units inside memory banks.

\subsection{Memory Systems for LLMs}
\label{subsec:background-hm-llm}
Figure~\ref{fig:hetero_system} illustrates four types of memory systems applicable to LLM inference acceleration.
The bandwidth-centric and the capacity-centric memory modules are represented as HBM and LPDDR, general solutions for modern AI acceleration~\cite{gracehopper, cxlpnm, hbm-pim}.

\noindent
\textbf{Multiple Bandwidth-Centric Memory:}
Modern GPU frameworks for LLMs typically configure a system resembling Figure~\ref{fig:hetero_system}~(a) by employing model parallelism across multiple GPUs, each equipped with tens of gigabytes of HBMs~\cite{orca, vllm}.
Nevertheless, such systems may encounter significant communication overhead from frequent synchronization among devices~\cite{deepspeed, petals}.
Moreover, although incorporating hundreds of gigabytes of HBM into a single module would satisfy both memory demands of capacity and bandwidth for LLMs, this approach is heavily constrained by cost factors and the nature of HBM architecture~\cite{hbm-arch, cxlpnm}.

\noindent
\textbf{Capacity-Centric Memory:}
Another feasible solution is depicted by Figure~\ref{fig:hetero_system}~(b), configuring the system entirely with capacity-centric memory.
This configuration provides scalability in capacity without incurring communication overhead.
CXL-PNM suggested employing cost-effective LPDDR5X memory due to its advantageous balance among memory bandwidth, capacity, and power consumption~\cite{cxlpnm}.
However, the capacity-centric memory can suffer from limited memory bandwidth compared to the bandwidth-centric memory.

\noindent
\textbf{Strict Hierarchical Memory:}
Traditional heterogeneous memory systems often assume a \textit{strict hierarchical memory} configuration, where computation units are exclusively attached to the bandwidth-centric memory, as illustrated in Figure~\ref{fig:hetero_system}~(c).
However, this approach requires data migration for every access to capacity-centric memory, resulting in power and latency overheads for workloads with limited locality~\cite{deepplan}.

\noindent
\textbf{Asymmetric Memory: }
To overcome the limitation of hierarchical memory, a heterogeneous memory system without a hierarchy between memory modules presents a viable solution.
As illustrated in Figure~\ref{fig:hetero_system}~(d), computation units (i.e. accelerators) are connected to both bandwidth-centric and capacity-centric memory in this system.
For example, NVIDIA's Grace Hopper Superchip and Grace Blackwell Superchip feature an architecture where a GPU with HBM and a CPU with LPDDR are connected via a high-speed interconnect~\cite{gracehopper, blackwell}.
By providing equal computational power to both memory devices, the system eliminates the necessity for frequent data migration between memory modules or direct access for computation.
In the following sections, we refer to this configuration as an \textit{asymmetric memory}.

\subsection{Host Memory Offloading Methods for LLMs}
\label{subsec:host-offloaded-system}

Several approaches have explored using host memory, a capacity-centric memory with relatively long access latency, for GPU-based ML computation~\cite{vdnn, cuda-um, deepplan, flexgen, infinigen, powerinfer}. DeepPlan uses a pipeline of load and execution, replacing some data migration with direct host-to-GPU memory access~\cite{deepplan}. FlexGen extends this by utilizing both the computational power and memory capacity of the host CPU~\cite{flexgen}.
While the large capacity of host CPU memory meets capacity demands, its low bandwidth and the limited bandwidth and high latency of PCIe interconnects significantly restrict overall performance. Consequently, system performance largely depends on which operation’s data is offloaded to the host memory. For instance, FlexGen determines the placement of weight parameters, KV cache, and activation vectors by solving a linear optimization problem for the system with GPU, CPU, and disk~\cite{flexgen}:

\vspace{-2.5ex}

\begin{equation}
\begin{array}{rrclcl}
\displaystyle \min_{p} & \multicolumn{3}{c}{Execution~Time} \\
\textrm{s.t.} 
&gpu\ peak\ memory &<& gpu\ mem\ capacity\\
&cpu\ peak\ memory &<& cpu\ mem\ capacity\\
&disk\ peak\ memory &<& disk\ mem\ capacity\\
&w_g+w_c+w_d&= &1\\
&c_g+c_c+c_d&= &1\\
&h_g+h_c+h_d&= &1
\end{array}
\label{eq:flexgen_linear}
\end{equation}



FlexGen addresses the challenge of efficiently utilizing host memory by optimizing the placement of data across GPU, CPU, and disk storage. The placement \(p\) is defined by nine variables: \((w_g, w_c, w_d)\) for weights, \((h_g, h_c, h_d)\) for activations, and \((c_g, c_c, c_d)\) for KV cache. These variables are relaxed to real values between 0 and 1 in the cost model, simplifying the optimization process. The problem is solved by adjusting these placement variables to minimize the objective function while adhering to memory constraints. This linear programming approach, outlined in Eq.~\eqref{eq:flexgen_linear}, balances memory usage and performance demands and can be extended to include additional considerations such as latency or compression techniques.

\section{Analysis of Mapping Space}
\label{sec:motiv}



\begin{figure}[t]
    \centering
    \includegraphics[width=0.95\columnwidth]{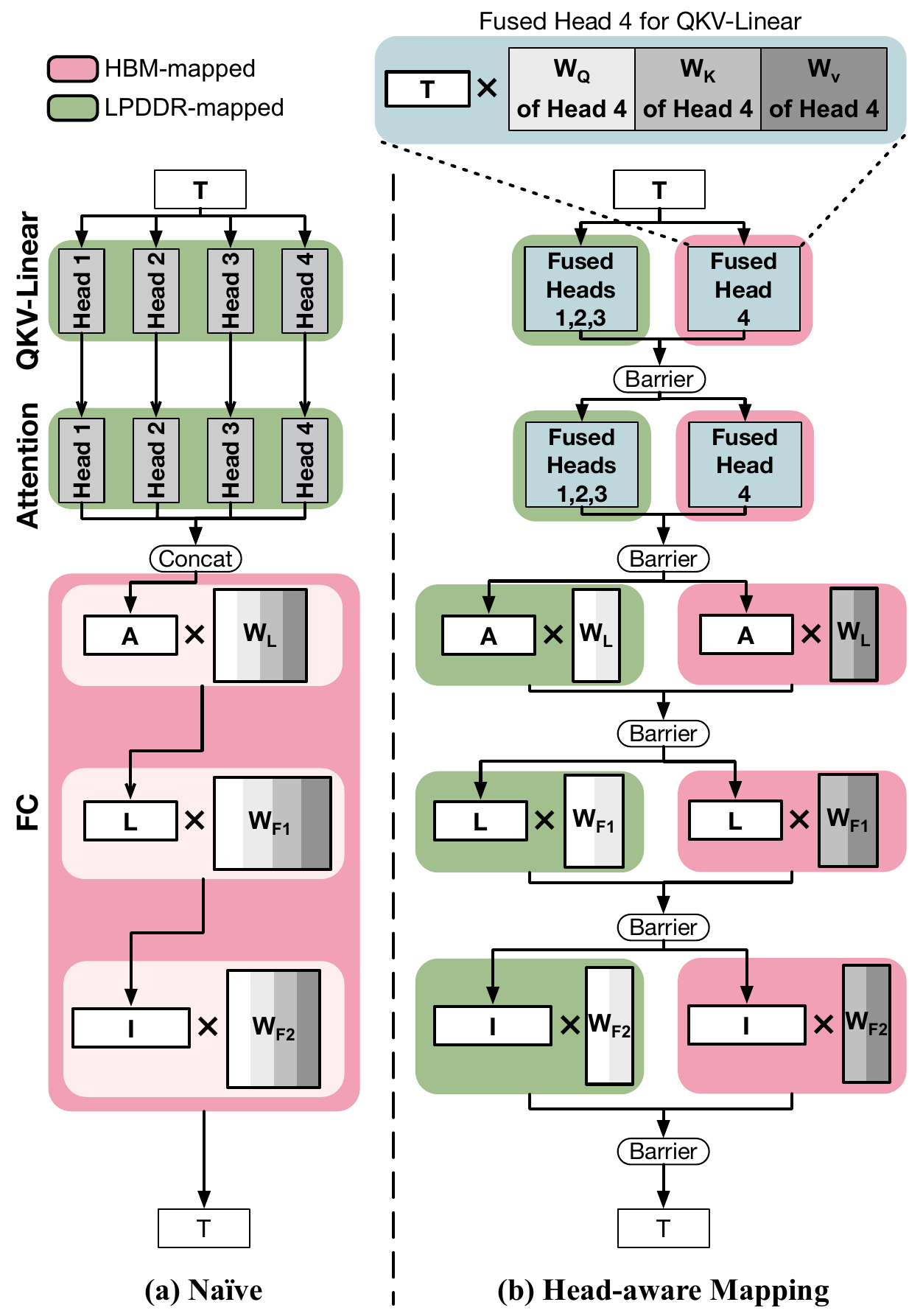}
    \vspace{-0.1in}
    \caption{Techniques for supporting optimal mapping granularity with asymmetric memory, for a single decoder layer.
    For clarity in explanation, a simplified view of a decoder layer is presented, visualizing GEMM/GEMV kernels mainly.}
    \vspace{-3ex}
    \label{fig:kernel-granularity}
\end{figure}

Serving LLMs imposes significant demands on bandwidth and memory capacity. Section~\ref{subsec:background-hm-llm} examines four potential strategies, three of which exhibit limitations in practicality or efficiency. To address these challenges, this paper proposes an asymmetric memory approach as an effective solution.

\subsection{Granularity of Kernel-Memory Mapping}
\label{sec:fine_grained_mapping}
To minimize redundant data migration, all LLM data should be mapped to asymmetric memory modules before each generation phase iteration.
Figure~\ref{fig:kernel-granularity}(a) shows a naive sublayer-granular mapping approach with model parallelism and KV cache parallelism, where decoder layers consist of \textit{qkv-linear}, \textit{attention}, and \textit{fc}.
As HBM-mapped operations must wait for LPDDR-mapped operations to complete due to strict dependencies, parallelization is not available and the accelerator cores are under-utilized.

To enhance LLM computation efficiency, various parallelization and kernel fusion techniques have been proposed
~\cite{megatron, deepspeed, seqparallelism1, seqparallelism2, lightseq, flashattention, flat, mamba, flexgen}.
Building on these techniques, we propose \textit{head-aware mapping granularity}, an advanced kernel-memory mapping granularity for asymmetric memory systems, as depicted in Figure~\ref{fig:kernel-granularity}(b).
In \textit{head-aware mapping granularity}, each sublayer is split into two parallel partitions and mapped to the HBM side and the LPDDR side, to maximize the utilization of both accelerator chips.

As shown in Figure~\ref{fig:kernel-granularity} (b), \textit{qkv-linear} and \textit{attention} are partitioned at the head granularity due to their independent heads. In contrast, the \textit{fc}, lacking a head-based structure, partitions GEMM kernels into two. To prevent partial sum accumulation, weight matrices are split column-wise, and activation matrices are fully copied into both memory modules. A synchronization barrier ensures computational correctness after each kernel.
Furthermore, kernels from multiple heads mapped to the same memory module can be fused into a single kernel call, as depicted in the upper part of Figure~\ref{fig:kernel-granularity}(b). This fusion increases matrix sizes for efficient blocked GEMM, improves on-chip memory utilization, and reduces kernel launch overhead.

\begin{figure}[t]
    \centering
    \includegraphics[width=\columnwidth]{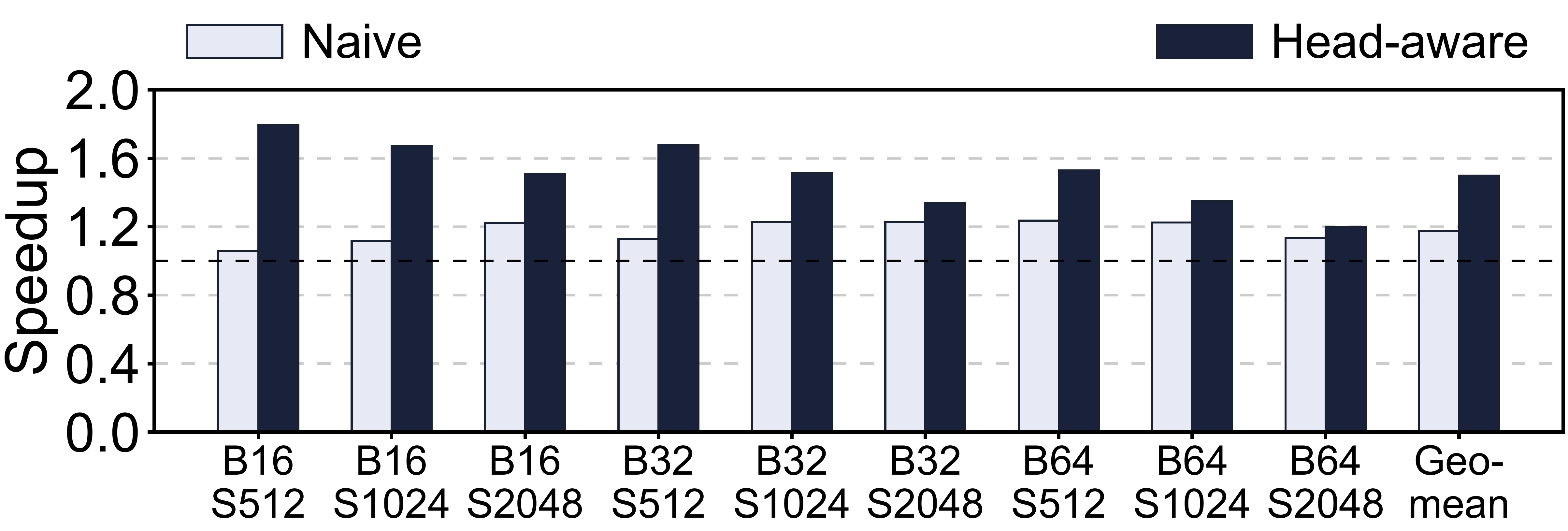}
    \vspace{-5ex}
    \caption{Comparison of the performance of the asymmetric memory with two mapping granularity options in GPT3-175B.}
    \label{fig:mapping-granularity-comparison}
    \vspace{-2ex}
\end{figure}

Figure~\ref{fig:mapping-granularity-comparison} compares the speedup of asymmetric memory with two mapping granularities, normalized to the baseline: LPDDR-only homogeneous memory (Figure~\ref{fig:hetero_system}(b)) with sublayer-granular mapping (Figure~\ref{fig:kernel-granularity}(a)). 
The x-axis represents batch size ($B$) and sequence length ($S$) (e.g., \textit{B16 S512} indicates a batch size of 16 and a sequence length of 512). 
The \textit{head-aware granularity} achieves a 1.50$\times$ speedup, outperforming the 1.27$\times$ speedup of the na\"ive granularity.
This underscores the effectiveness of the proposed \textit{head-aware mapping granularity}, which is adopted as a foundational assumption for all asymmetric memory variants discussed in the paper.
Note that the performance of both options is evaluated using their optimal kernel-memory mapping, defined as the configuration yielding the best performance. 


The key concepts of the proposed head-aware mapping granularity are (1) overlapping the execution time of two sides, (2) reducing the capacity consumption by distributing model parameters and KV cache exclusively, and (3) reducing the synchronization overhead by exploiting the parallelism of heads.
Therefore, the head-aware mapping granularity is applicable to other variants of LLMs as long as there exists an independency between kernels inside a sublayer for most sublayers (e.g. `head' and `expert' of mixture-of-expert (MoE) models).

\subsection{Limitation of Mapping Policy from Host Memory Offloading}
\label{subsec:vs-flexgen}
Another important feature that makes significant impact on the LLM inference performance in asymmetric memory is, \textit{kernel-memory mapping} decision.
That is, the number and type of kernels mapped to each side of the asymmetric memory decides the overall performance of the system.
Prior work, FlexGen proposed an analytical performance model-based mapping strategy to guarantee performance in systems with host memory offloading~\cite{flexgen}, which is described in Equation~\ref{eq:flexgen_linear}.

\begin{figure}[t]
    \centering
    \includegraphics[width=\columnwidth]{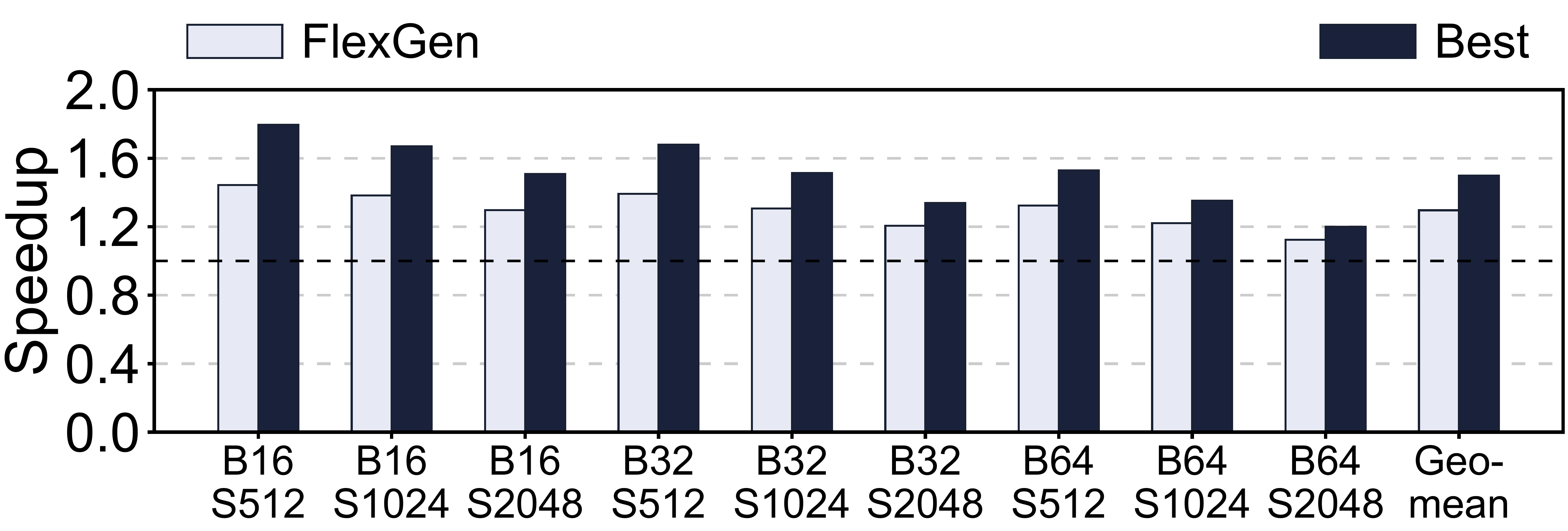}
    \vspace{-5ex}
    \caption{The relative speedup over the baseline with two mapping policies of asymmetric memory in GPT3-175B, batch size 32. \texttt{FlexGen} follows the mapping found by the model of Equation~\ref{eq:flexgen_linear}, and \texttt{Best} follows the mapping decision found by exhaustive search with $N^3$ times profiling.}
    \label{fig:hetero-mapping-vs-flexgen}
    \vspace{-2ex}
\end{figure}

Figure~\ref{fig:hetero-mapping-vs-flexgen} compares two mapping strategies: Mapping decision from FlexGen's performance model (Equation~\ref{eq:flexgen_linear}, denoted as \texttt{FlexGen}), and the optimal mapping determined through $N^3$ times profiling and exhaustive search (\texttt{Best}).
The x axis represents the batch size and sequence length pair, and the y axis is a normalized speedup over the baseline.
We modified FlexGen's performance model illustrated in Equation~\ref{eq:flexgen_linear} to suit asymmetric memory systems instead of single-GPU systems.
As shown in Figure~\ref{fig:hetero-mapping-vs-flexgen}, while \texttt{Best} shows 1.50$\times$ speedup over the baseline on average, \texttt{FlexGen} reports 1.30$\times$ speedup on average - 0.87$\times$ of \texttt{Best}.

The mapping decisions of the FlexGen model are suboptimal in the asymmetric memory system for two key reasons. First, since FlexGen is designed for offline inference, it does not account for changes in batch size or sequence length, leading to notable performance degradation in our asymmetric memory system due to static mapping.
Second, its performance model fails to account for the distinct characteristics of the three sublayer types.
As explained in Section~\ref{subsec:host-offloaded-system}, FlexGen model's partitioning scheme is based on three groups: Weight, activation, and KV cache.
Therefore, \textit{qkv-linear} and \textit{fc} are considered as a same group, partitioned within the same ratio as well as ignoring their different characteristics such as dimensions of weight matrices, number of GEMM kernels, parallelism between heads, and existence of the barrier.
Moreover, as the FlexGen model only considers the total capacity and FLOP assigned to each side of memory device, the high memory intensiveness of \textit{attention} is ignored.
Because of this, in particular, mapping \textit{attention} to LPDDR causes significant performance degradation, as the slowdown of memory-intensive GEMV kernels in \textit{attention} affects the entire system.
In addition, the mapping of \texttt{FlexGen} remains static when batch sizes and sequence lengths change, which might be efficient for the offline inference that FlexGen was originally designed for, but not efficient for the scenario with dynamic changes of batch size and sequence lengths.
Note that the growth of batch size and sequence length incurs the narrower gap between \texttt{FlexGen} and \texttt{Best} due to the decrease of relative portion of HBM in total footprint: As the relative portion of HBM in total memory consumption of the system gets smaller, the performance of the systems with both mapping gets closer to the baseline (i.e. LPDDR-only system), resulting in the smaller gap between two mapping strategies.

Although using FlexGen’s performance model to determine a mapping for asymmetric memory is unsuitable, finding the optimal mapping through exhaustive search is also impractical due to its high profiling cost. Assuming the number of attention heads is $N$, the search space is $N^3$ (\textit{qkv-linear} $\times$ \textit{attention} $\times$ \textit{fc} decisions), amounting to 884,736 options for the GPT3-175B model with 96 heads. Profiling such a vast space for each LLM inference iteration introduces significant overhead, emphasizing the need for a new mapping policy. In the following, we quantitatively analyze the correlation between kernel-memory mapping decisions and performance.

\begin{figure}[t]
    \centering
    \includegraphics[width=\columnwidth]{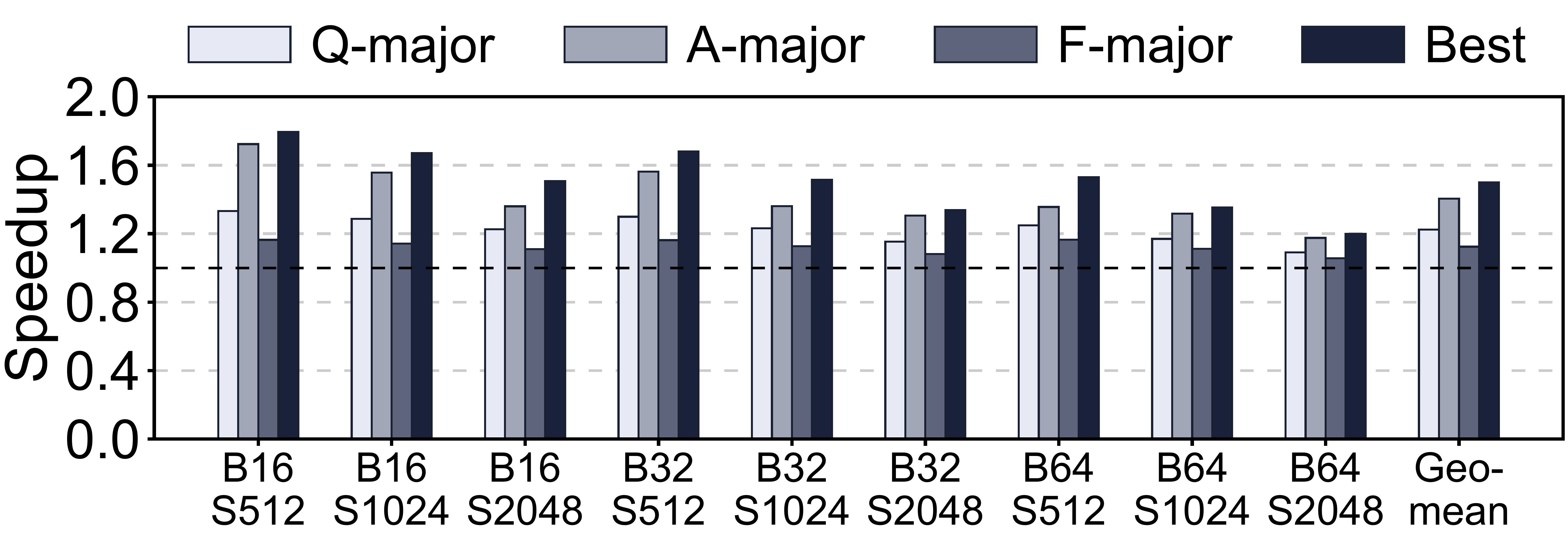}
    \vspace{-5ex}
    \caption{The relative speedup over the baseline for asymmetric memory variants in GPT3-175B, batch size 32. \texttt{Q-major}, \texttt{A-major} and \texttt{F-major} follows the mapping decision found by exhaustive search among result of $N^2$ times profiling, each favoring HBM for \textit{qkv-linear}, \textit{attention}, and \textit{fc}, respectively. \texttt{Best} follows the mapping decision found by exhaustive search with $N^3$ times profiling.}
    \label{fig:hetero-mapping-vs-firsttouch}
    \vspace{-2ex}
\end{figure}

\begin{figure*}[t]
    \centering
    \includegraphics[width=\textwidth]{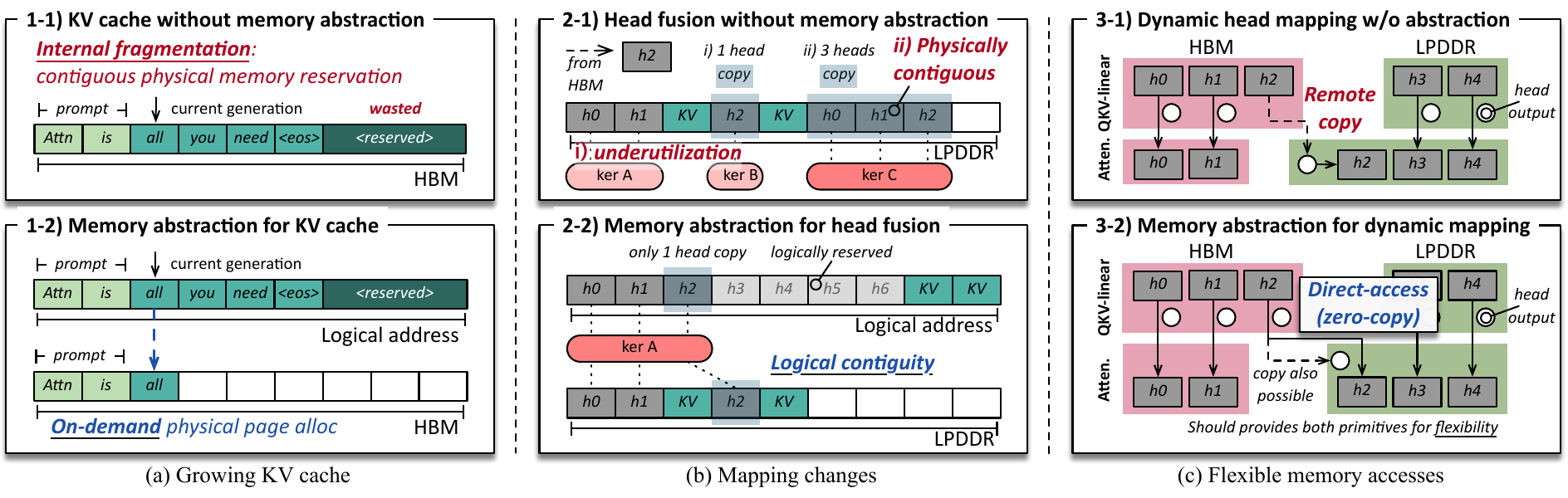}
    \vspace{-5.0ex}
    \caption{The need for dynamic memory management in asymmetric memory systems: three motivations for memory abstraction to decouple the logical address space from the physical address space.}
    \label{fig:arch-memory-layout}
    \vspace{-1ex}
\end{figure*}

\subsection{Importance of Sublayer Characteristic Consideration}
\label{subsec:tendency}
Figure~\ref{fig:hetero-mapping-vs-firsttouch} compares the performance of the best mapping (\texttt{Best}) with three alternative mappings with smaller search space.
Each of \texttt{Q-major}, \texttt{A-major}, and \texttt{F-major} means an alternative best mapping decision with keeping as much \textit{qkv-linear}, \textit{attention}, and \textit{fc} in HBM as possible, respectively (i.e. Requiring $N^2$ profiling followed by exhaustive search each).
The performance is measured as a speedup over the baseline, for each batch size and sequence pair.
While \texttt{Best} shows the performance 1.50$\times$ faster than the baseline, \texttt{Q-major} and \texttt{F-major} only reaches 1.22$\times$ and 1.12$\times$ speedup over the baseline, respectively.
On the other hand, \texttt{A-major} shows notable performance improvement compared to other two options: 1.40$\times$ speedup over the baseline, 0.94$\times$ of \texttt{Best}.
%
Such difference is originated by distinct characteristics of three sub-layers:
Each sublayer of LLM have a notable difference in the correlation between the available memory bandwidth and their performance.

\noindent
\textbf{Importance of Attention Sublayer Mapping:}
Since GEMV kernels of \textit{attention} makes it highly bandwidth-bounded, the performance of \textit{attention} with HBM shows higher performance boost over LPDDR than other two sublayers.
In addition, the impact of \textit{attention} on overall performance is proportional to the batch size and sequence length, as these values are directly related to the size of KV cache.
This is the reason why \texttt{A-major} consistently showed good performance in Figure~\ref{fig:hetero-mapping-vs-firsttouch}.

\noindent
\textbf{Other Sublayers:}
%
%
Compared to the \textit{attention} layer, \textit{qkv-linear} and \textit{fc} share similar characteristics. While the \textit{attention} layer maintains a nearly constant arithmetic intensity, the arithmetic intensity of \textit{qkv-linear} and \textit{fc} increases with the batch size. Additionally, their weight parameters and activations remain stable regardless of sequence length, and their overall footprint varies minimally with batch size, leading to a relatively consistent memory demand.
As the LLM footprint grows due to KV cache expansion, the contribution of \textit{qkv-linear} and \textit{fc} to overall performance decreases. This trend explains the consistently suboptimal performance of \texttt{Q-major} and \texttt{F-major}, as shown in Figure~\ref{fig:hetero-mapping-vs-firsttouch}. These characteristics emphasize the need for carefully tailored kernel-memory mapping strategies for each sublayer type.

\subsection{Need for Dynamic Memory Management}
Along with the unique characteristics of LLM operations and fine-grained kernel-memory mapping, the distinct traits of memory modules in an asymmetric memory introduce significant challenges for deploying LLMs.
Efficient LLM execution requires dynamic memory management to handle growing KV caches, adapt to runtime changes in kernel-memory mapping, and optimize data placement.
Figure~\ref{fig:arch-memory-layout} presents three common challenges in using asymmetric memory for LLMs and demonstrates how our dynamic memory management resolves them. To avoid underutilization and migration overhead from direct physical memory allocation, we propose decoupling logical and physical address spaces. This approach enables flexible memory access and abstraction, ensuring effective use of both memory modules and accelerators, ultimately maximizing LLM performance on asymmetric memory systems.

\noindent
{\bf (a) Growing KV cache: } Since the KV cache grows with more generated tokens, using physical memory addresses directly requires pre-allocating contiguous memory to accommodate the maximum sequence length for each request. However, as shown in Figure~\ref{fig:arch-memory-layout} (1-1), much of this reserved memory is wasted, as not all requests reach the maximum sequence length. This inefficiency severely impacts asymmetric memory systems, where HBM’s limited memory space is crucial for performance. vLLM previously addressed this issue by introducing a software-based virtualization layer for the KV cache, allowing more flexible memory allocation~\cite{vllm}.

Building on this concept, as shown in Figure~\ref{fig:arch-memory-layout} (1-2), we propose a hardware-integrated memory virtualization approach tailored for asymmetric memory systems. Unlike vLLM’s software-based solution, our method dynamically manages memory at the hardware level, enabling the KV cache to be allocated contiguously in logical address space while assigning physical memory only as needed for generated tokens~\cite{vattention}. This eliminates the complexity of software-based translation while optimizing memory utilization.
Moreover, our method is applicable for all tensors and all sublayers of LLM, while PagedAttention approach of vLLM is designed only for KV cache (i.e. only applicable for \textit{attention} sublayer).

\noindent
{\bf (b) Mapping changes: } In asymmetric memory systems with the fine-grained kernel-memory mapping proposed in Section~\ref{sec:fine_grained_mapping}, the best mapping may change at runtime. 
For instance, a tensor composed of three heads (\textit{h0}, \textit{h1}, \textit{h2}) might initially be partitioned between LPDDR (\textit{h0}, \textit{h1}) and HBM (\textit{h2}).
Figure~\ref{fig:arch-memory-layout} (2-1) describes a scenario where the entire tensor needs to be stored in LPDDR due to a runtime change in the best mapping, requiring \textit{h2} to be moved from HBM to LPDDR.
In such cases, the tensor must be stored in a contiguous region in LPDDR to prevent performance degradation caused by dividing a matrix operation into multiple kernels.

Without memory abstraction, transferring \textit{h2} requires allocating a new physically contiguous region and copying all three heads to the new location.
Such data movement leads to significant under-utilization of LPDDR, as each head is stored in LPDDR twice.
As illustrated in Figure~\ref{fig:arch-memory-layout} (2-2), we address this under-utilization problem by dynamically managing asymmetric memory with memory abstraction.
The proposed approach enables each kernel to access data through a contiguous logical memory region, even when physical pages are scattered across the memory.

\noindent
{\bf (c) Flexible memory accesses: } 
When using fine-grained mapping in asymmetric memory systems, the output of a layer may be computed across accelerators attached to two different memory modules. 
For the next layer, each accelerator may require access to the output stored in the other side.
Figure~\ref{fig:arch-memory-layout} (3) describes two scenarios where the accelerator on the LPDDR side requires \textit{h2}, which is stored on the HBM side.
Depending on the degree of data reuse, the data can either be copied to a different memory module (Figure~\ref{fig:arch-memory-layout} (3-1)) or direct-accessed remotely (Figure~\ref{fig:arch-memory-layout} (3-2)). 
To enable such dynamic memory management, memory abstraction provides a flexible memory access mechanism, allowing kernels to operate within a logical address space without requiring modifications to the kernel code.

\section{Design}
\label{sec:arch}

\subsection{Overview}
\label{subsec:arch-overview}
\begin{figure}[t]
    \centering
    \includegraphics[width=\columnwidth]{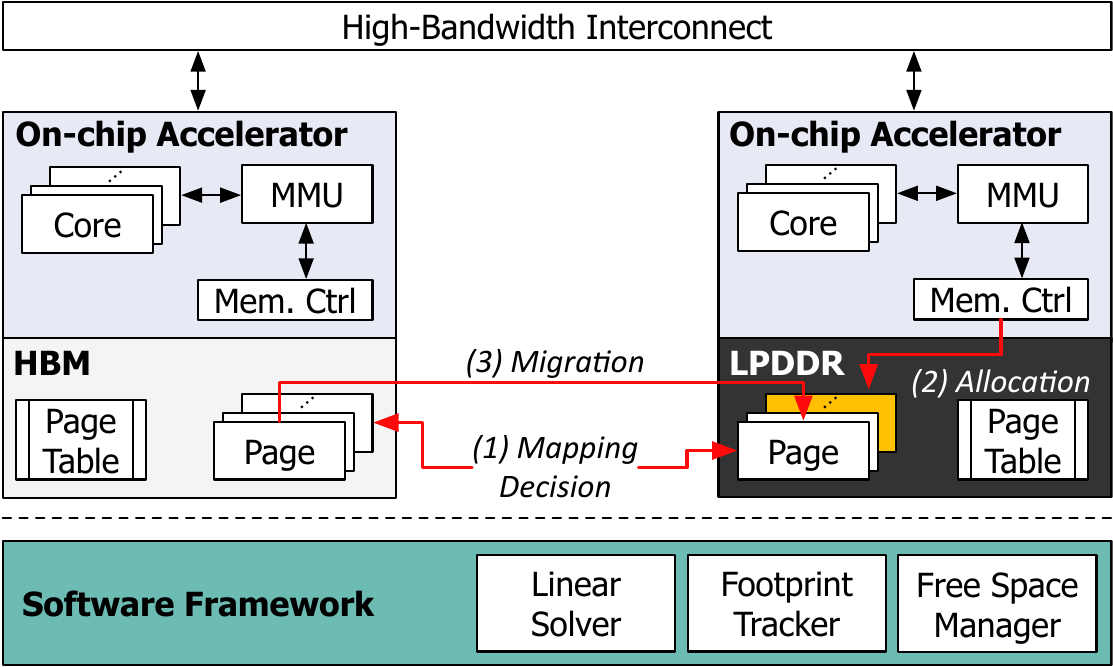}
    \vspace{-2ex}
    \caption{Overview of \amalgam with hardware substrates.}
    \label{fig:hw-support}
    \vspace{-1ex}
\end{figure}

This paper proposes a dynamic mapping algorithm and a hardware support for memory abstraction to maximize the potential of asymmetric memory for LLM processing.
Figure~\ref{fig:hw-support} shows the overall organization of our system \abcmem.

\begin{figure}[t]
    \centering
    \includegraphics[width=\columnwidth]{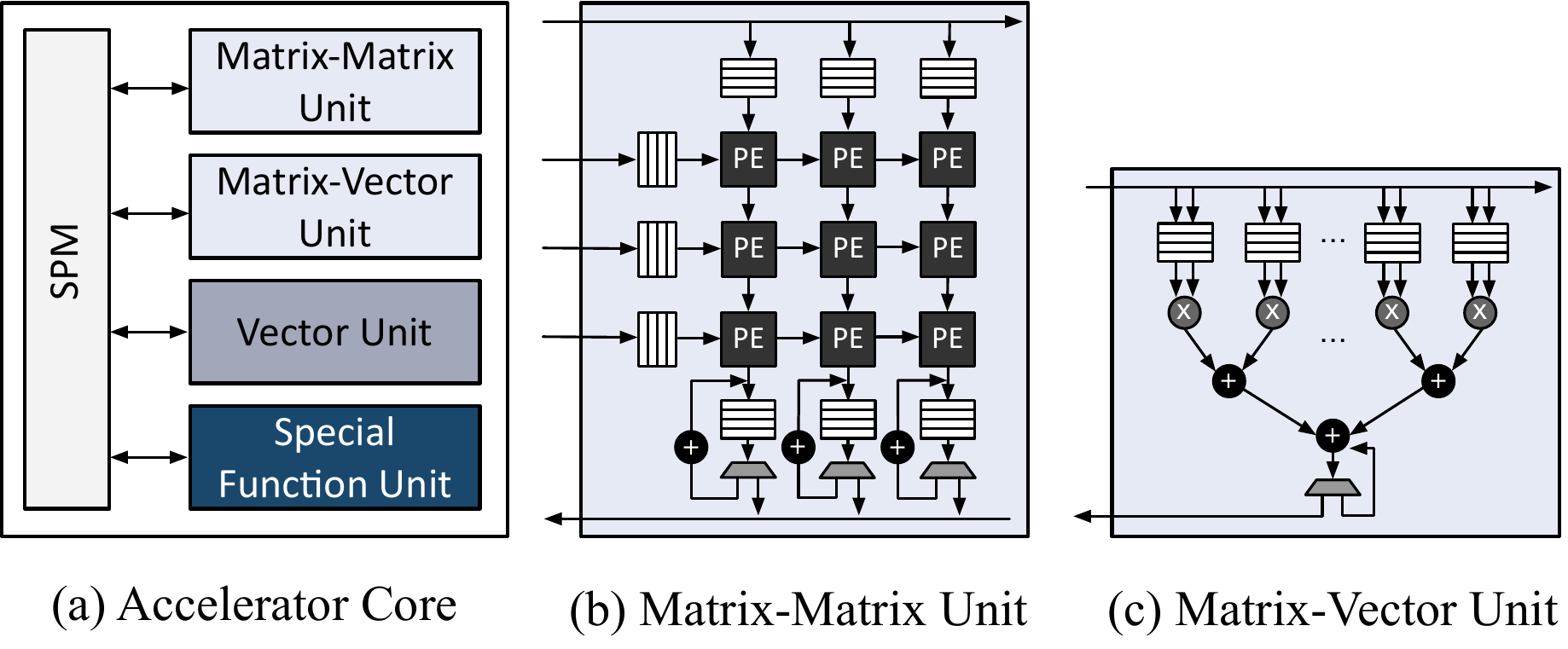}
    \vspace{-4ex}
    \caption{Architecture of the accelerator core: (a) Overview, (b) Matrix-Matrix multiplication unit, and (c) Matrix-Vector multiplication unit. An accelerator chip of each memory module includes 4 cores.}
    \label{fig:core-config}
    \vspace{-2ex}
\end{figure}

\noindent
\textbf{Hardware Substrate:}
In our system, the asymmetric memory consists of two different memory modules: bandwidth-centric HBM3 with 3TB/s bandwidth, and capacity-centric LPDDR5X with 544GB/s bandwidth~\cite{gracehopper}. 
On each side of the asymmetric memory, there is an accelerator chip processing kernel computations of LLMs.
Each accelerator chip is comprised of four cores, and all accelerator units in each core share the on-chip scratchpad memory (SPM) with double buffering mechanism. The memory modules and accelerators are placed in a single board with PCIe interconnect to the host.
Figure~\ref{fig:core-config} illustrates the architecture of each accelerator chip. Considering various types of operations in LLM, the accelerator core contains four types of accelerator units:
matrix-matrix (MM) unit for GEMMs, matrix-vector (MV) unit for GEMVs, vector unit for layer normalization and residual connections, and special function unit (SFU) for activation functions.
Section~\ref{subsec:method} explains the detailed hardware parameters.

\noindent
{\bf Memory abstraction support: } Instead of using physical memory directly, \abcmem provides memory abstraction to address the KV cache allocation problem and the dynamically changing kernel-memory mapping (Section~\ref{subsec:memory-abstraction}). 
Each accelerator chip in \abcmem includes an MMU for address translation to support memory abstraction. 
A simple flat page table for each side is maintained by the host driver. 
The page tables in the HBM and LPDDR sides may have different contents, as the same virtual page containing weights can be duplicated in both HBM and LPDDR.

\noindent
{\bf Dynamic mapping support: } The memory mapping can change for two major reasons. First, the KV cache can grow with increasing sequence lengths, requiring the addition of a new physical page. Second, a request can complete earlier than the other requests in the same batch, prompting the addition of a new request to the batch. This can shrink or expand the KV cache size, leading to a change in the mapping decision (Section~\ref{subsec:remapping-decision}). 
When such a mapping change occurs, the host driver updates both page tables and invalidates the TLBs in the MMUs. 

\noindent
{\bf Kernel synchronization: } As shown in Figure~\ref{fig:kernel-granularity}, kernels executed in two memory sides often need to be synchronized before advancing to the next operation.
To optimize synchronization efficiency while reducing kernel launch overhead, we adopt a hardware-based synchronization mechanism similar to the CUDA event for NVIDIA GPUs \cite{cuda-event}.
The \abcmem driver initiates all necessary LLM kernels on both accelerators, managing dependencies via the \abcmem hardware controller rather than explicit host-side synchronization.
Kernels register their dependencies upon launch and are temporarily removed from the scheduling pool, to be reintroduced when their prerequisites are complete.
This approach minimizes kernel launch overhead and accelerator idle time, thereby enhancing overall synchronization efficiency.

\subsection{Dynamic Memory Management}
\label{subsec:memory-abstraction}

\subsubsection{Page-based Memory Virtualization}
We apply traditional page-based virtualization for the asymmetric memory system, to support dynamic memory management through memory abstraction.
Two page tables are maintained for the HBM and LPDDR sides, which are maintained by the host driver. 
Each accelerator has an MMU with 2048 TLB entries, referring to the prior work~\cite{neummu, mnpusim}.
To reduce the complexity and latency of page table walks, we use a simple flat page table. 
Unlike conventional CPUs, the logical address space is not huge for the LLM accelerator, as the entire capacity is bounded by the sum of HBM and LPDDR capacity plus the extra KV cache space for maximum sequence length. 
Therefore, instead of employing a radix-tree page table, a flat table allows a single memory access for a TLB miss. 
Assuming 1TB of logical address space and 2MB page size, the page table size is only 4MB.

\noindent
{\bf Page Size: }
\label{subsubsec:pagesize}
Considering the overhead of address translation, our system employs a \textbf{2MB huge page} for our system.
However, the use of large page sizes may present a risk of internal fragmentation.
Hence, we investigate the potential data fragmentation using GPT3-175B with a batch size of 32, to estimate the impact of internal fragmentation.
The size of internal fragmentation for each type of tensor is calculated using Equation~\ref{eq:fragmentation-size}.

\vspace{-1.5ex}
\begin{equation}
    \small
    \begin{array}{l}
        \textit{fragmentation} = ((\textit{tensor\_size})~mod~(\textit{page\_size}))\times(\textit{\# tensors})\\
    \end{array}
    \label{eq:fragmentation-size}
\end{equation}

The term \textit{tensor\_size} denotes the minimum size of consecutive data in each tensor, which is consistently mapped to the same memory module (either bandwidth-centric or capacity-centric) and cannot be merged with other tensors.
By accumulating the sizes of internal fragmentation calculated using Equation~\ref{eq:fragmentation-size}, we find that the total maximum internal fragmentation amounts to 156MB for GPT3-175B, which occupy only 0.16\% of the HBM capacity.
Based on the analysis, a 2MB page size is sufficient for current system and target workloads.

\subsubsection{HW/SW Support for the Change of Mapping}
While the batch size and sequence length of LLMs can change at runtime, such changes appear only at the end of each iteration in the generation phase, meaning they occur infrequently.
As illustrated in Figure~\ref{fig:hw-support}, three types of events can occur at the end of each iteration: (1) mapping decision, (2) allocation, and (3) migration.
These events are executed under the control or assistance of software frameworks (i.e., \textit{linear solver}, \textit{footprint tracker}, \textit{free space manager}) with support from the hardware components described in the hardware substrate.
Note that due to the hardware support and relatively long intervals between batch size or sequence length update, these events have minimal impact on overall system performance.

\noindent
\textbf{Mapping Decision:}
\textit{Linear solver} is the software implementation of the mapping algorithm explained in Section~\ref{subsec:remapping-decision}.
The mapping decision is updated when the \textit{linear solver} determines that a change is required based on the information from the \textit{footprint tracker}.
In such cases, the subsequent allocation and migration processes are triggered, resulting in changes to the kernel-memory mapping.

\noindent
\textbf{Allocation:}
Physical page allocation is required for newly generated tokens or additional requests.
The location of newly allocated physical pages is determined by the \textit{free space manager}, and the allocation mechanism also updates the information of MMUs.

\noindent
\textbf{Migration:}
When the mapping changes, pages previously mapped to HBM need to be migrated to LPDDR, or vice versa.
In this process, the MMU's information is updated accordingly, and
the allocation mechanism is applied if a new page allocation is required at the destination.

\subsection{Runtime Dynamic Mapping Decision}
\label{subsec:remapping-decision}

\begin{algorithm}[t]
\caption{Mapping Decision Algorithm}
\label{algorithm:initial_mapping}
\begin{algorithmic}[1]
    \State \textbf{def} initialize\_mapping($N$: Number of heads):
        \State \hskip1.5em \textbf{for} sublayer \textbf{in} [\textit{attention}, \textit{qkv-linear}, \textit{fc}]:
            \State \hskip3.0em Find $n$ such that 
                \State \hskip3.0em $\bullet$ Possible to map $n$ heads of the sublayer to HBM
                \State \hskip3.0em $\bullet$ Possible to map $(N-n)$ of the sublayer to LPDDR
                \State \hskip3.0em $\bullet$ Minimize the peak execution time on both 
                \State \hskip3.75em HBM and LPDDR sides
\end{algorithmic}
\end{algorithm}

\subsubsection{Requirements of Mapping Decision Strategy}
As explained in Section~\ref{sec:motiv}, an effective kernel-memory mapping can significantly boost performance in asymmetric memory systems.
However, finding the \textit{best-mapping} through exhaustive search and profiling is impractical, as head-aware mapping in asymmetric memory involves $N$$\times$$N$$\times$$N$ choices, where $N$ is the number of heads per each decoder.
To maximize performance in asymmetric memory, the mapping strategy must satisfy three key conditions.
First, the mapping decision should consider the \textbf{runtime change of performance trend} caused by changes in batch size and the sequence length.
Second, the mapping decision should \textbf{reflect the characteristics of each sublayer}.
As discussed in Section~\ref{subsec:tendency}, differences in arithmetic intensity and memory footprint imply that the impact of the mapping decision on overall performance varies across sublayers.
Lastly, the \textbf{cost} of the mapping decision process should be \textbf{low}.
Even if a mapping decision is optimal, 
excessive decision-making costs (e.g., requiring extensive profiling) can negate the performance gains by introducing significant overhead.


\subsubsection{Mapping Algorithm}
\label{subsubsec:decision-alg}
Algorithm~\ref{algorithm:initial_mapping} shows our mechanism for runtime dynamic mapping decision, satisfying all three conditions explained above.
Algorithm~\ref{algorithm:initial_mapping} aims to maximize the performance of both HBM and LPDDR sides without encountering out-of-memory issues.
Since \textit{attention} requires more HBM capacity than \textit{qkv-linear}, we prioritize mapping optimization in the order of \textit{attention}, \textit{qkv-linear}, and \textit{fc} as shown in line 2.
The first and second bullet of line 3 aim to prevent out-of-memory issues.
For the third bullet of line 3, we employ the min-max algorithm to ensure balanced execution times across both memory sides for each sublayer.
To reflect the characteristics of each sublayer, the peak execution model first calculates the \textit{ideal execution time} by dividing the total number of arithmetic operations by the maximum throughput of the accelerator chip.
Next, a hyperparameter that reflects the arithmetic intensity is multiplied to the \textit{ideal execution time} for additional elaboration.
The cost of solving Algorithm~\ref{algorithm:initial_mapping} with single-threaded C++ implementation is 0.05ms with Intel i7-6700 CPU, which is nearly negligible compared to the cost of LLM inference.
Compared to the exhaustive search-based mapping strategies (i.e. mappings of Figure~\ref{fig:hetero-mapping-vs-firsttouch}, our mapping algorithm does not requires profiling in $O(N^2)$ or $O(N^3)$ space.
Instead, it only needs to solve simple polynomials to find the nearly-optimal mapping, which leads to the low temporal overheads.

Basically, Algorithm~\ref{algorithm:initial_mapping} operates as a greedy algorithm that prioritizes HBM allocation in the order of \textit{attention}, \textit{qkv-linear}, and \textit{fc} sublayers.
Even if the batch size and sequence length change at runtime, the mapping decisions determined by this algorithm remain relatively stable, as long as the changes do not cause significant change in HBM utilization.
Therefore, even if a new (i.e. changed) mapping decision necessitates altering the existing mapping, the amount of data migration required is relatively small.
For instance, in typical LLM inference scenarios where sequence length continually increases with token generation, eviction from HBM to LPDDR occurs in order of \textit{fc}, \textit{qkv-linear}, and \textit{attention}.
By following the Algorithm~\ref{algorithm:initial_mapping}, once a layer is evicted, there is no need to bring it back into HBM in this scenario.
This minimizes a redundant data migration during runtime.

\section{Evaluation}
\label{sec:eval}

\begin{table}[t]
    \centering
    \footnotesize
    \begin{tabular}{ccc}
        \toprule
        \textbf{Arch} & \textbf{Capacity} & \textbf{Bandwidth} \\ 
        \midrule
        HBM3 & 96GB & 3TB/s \\ 
        LPDDR5X & 512GB & 544GB/s \\ 
        Interconnect & - & 960GB/s \\
        \toprule
    \end{tabular}
    \caption{Heterogeneous memory system configuration, following the configuration of ~\cite{gracehopper}.}
    \label{tab:memconfig}
    \vspace{-5ex}
\end{table}

\begin{table}[t]
    \centering
    \footnotesize
    \begin{tabular}{cc}
        \toprule
        \multicolumn{2}{c}{\textbf{MV Unit}} \\
        \toprule
        PE Configuration & 32 $\times$ 1D Array (128 $\times$ 1) \\
        Algorithm & Dot product \\
        \toprule
        \multicolumn{2}{c}{\textbf{MM Unit}} \\
        \toprule
        PE Configuration & Systolic Array (128 $\times$ 128) \\
        Dataflow & Weight stationary \\
        \toprule
        \multicolumn{2}{c}{\textbf{SFU \& Vector Unit}} \\
        \toprule
        ADD/SUB/MUL/DIV & 1D Array (128 $\times$ 1) \\
        Adder Tree & 128 Adders \\
        Lookup Table & 128 req/cycle\\
        \toprule
        \multicolumn{2}{c}{\textbf{Common}} \\
        \toprule
        Core Frequency & 1GHz \\
        On-chip SPM Size & (16MB $\times$ 2) per core \\
        HBM Access Latency & 32ns \\
        LPDDR Access Latency & 45ns \\
        TLB Miss Latency & 300ns \\
        \toprule
    \end{tabular}
    \caption{Accelerator parameter configuration.}
    \label{tab:archconfig}
    \vspace{-5ex}
\end{table}

\subsection{Methodology}
\label{subsec:method}
\noindent
\textbf{Architecture Configuration:}
Table~\ref{tab:memconfig} shows the major parameters of two memory modules.
The interconnect refers to the connection between accelerators in two memory modules. We adopt the memory configuration of NVIDIA Grace Hopper Superchip for the heterogeneous memory configuration~\cite{gracehopper}.
Table~\ref{tab:archconfig} defines the detailed hardware configuration of the accelerator units.
We adopt the systolic array configuration of Google Cloud TPU for the MM units~\cite{cloudtpu}.
We refer to the vector unit configuration of DFX~\cite{dfx}, but scale up the number of computation units considering the size of the systolic array.

\noindent
\textbf{Baseline:}
We adopt the capacity-centric memory system in the prior work as a baseline~\cite{cxlpnm}.
All comparison configurations, including the baseline system configuration, consist of two accelerator chips that provide the same computational power with an asymmetric memory system.

\noindent
\textbf{Simulation:}
We model homogeneous and heterogeneous memory systems by developing a cycle-level performance simulator, cross-validated by profiling the open-source multi-core NPU simulator and DRAM simulator~\cite{mnpusim, dramsim3}.
We apply the parameters defined in Table~\ref{tab:memconfig} and ~\ref{tab:archconfig} for the simulation.
We collect memory access latency data from Ghose et al.~\cite{dram-analysis}.

\noindent
\textbf{Benchmarks:}
We use three modern decoder-based LLMs: GPT3, Chinchilla and Llama2 as our evaluation benchmark~\cite{gpt3, chinchilla, llama2}.
Among variants of these models, we select GPT3-175B with 175 billion parameters, Chinchilla-70B and Llama2-70B with 70 billion parameters each to model the scenario with insufficiency of HBM capacity.
Note that we select Llama2-70B in addition to Chinchilla-70B to analyze the effect of grouped-query attention (GQA), despite their similar model parameter sizes.
We assume INT8 precision for all three models, as we use ASIC accelerator chips to provide the computational power.

\noindent
\textbf{Performance Measurement:}
We only include decoder layers in our evaluation, as decoder layers occupy a significant portion of computations in decoder-based LLMs.
Additionally, referring to the prior work, we measure the performance for a single iteration of the generation phase with a given batch size and sequence length~\cite{neupims}.
In this section, we fix the batch size for each model: 32 for GPT3, 64 for Chinchilla, and 128 for Llama2.

\noindent
\textbf{Evaluation Metric:}
We use the relative speedup as a main metric of performance evaluation, which is defined as (time elapsed for a single iteration of the baseline)/(time elapsed for a single iteration of \amalgam).
The relative speedup has same meaning with relative throughput and relative latency improvement in our evaluation scenarios:
First, as a relative throughput of) \amalgam can be calculated as (the throughput of \amalgam)/(the throughput of the baseline) and the throughput is equal to (batch size)/(time elapsed for a single iteration), the relative throughput has the same value with the relative speedup.
Similarly, the latency improvement for time-between-token (TBT) can be calculated by (time elapsed for a single iteration of the baseline)/(time elapsed for a single iteration of \amalgam), sharing a same definition with the relative speedup.
Note that we do not include time-to-first-token (TTFT) metric in our evaluation, as TTFT is a latency evaluation metric for prompt phase, which is out of our scope.


\subsection{Speedup}
\label{subsec:speedup}
\subsubsection{Effectiveness of Asymmetric Memory}
We first compare the performance of \amalgam with two alternative configurations:
A strict hierarchical memory architecture (\texttt{Hierarchical}) and an asymmetric memory architecture always using the best kernel-memory mapping (\texttt{Oracle}).
Note that \texttt{Oracle} assumes no overhead from memory abstraction, by setting the cost of PTW/TLB access as zero.

\begin{figure}[t]
    \centering
    \includegraphics[width=\columnwidth]{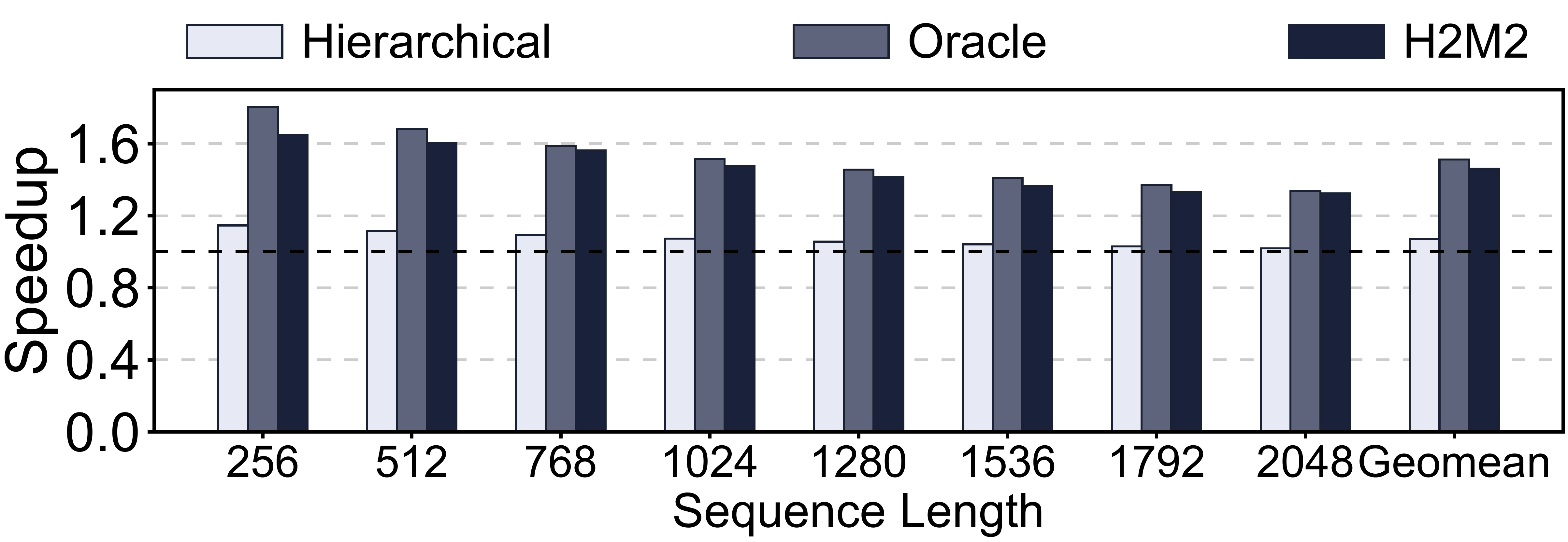}
    \vspace{-4ex}
    \caption{The relative speedup over the baseline (LPDDR-only) for hierarchical memory and asymmetric memory in GPT3-175B, batch size 32.}
    \label{fig:eval-hetero-gpt3}
    \vspace{-2ex}
\end{figure}

\noindent
\textbf{GPT3-175B:}
Figure~\ref{fig:eval-hetero-gpt3} visualizes the performance of GPT3-175B with the four configurations, translated to the relative speedup over the baseline LPDDR-only system.
While the strict hierarchical memory shows a 1.07$\times$ speedup over the baseline on average, \amalgam reports a 1.46$\times$ speedup on average, reaching 0.97$\times$ of the ideal asymmetric memory performance.
Due to the lack of temporal locality in LLM, on-demand migration of \texttt{Hierarchical} yields migration overhead, leading to relatively low performance boost.
On the other hand, \amalgam takes an advantage of heterogeneous memory through nearly-optimal kernel-memory mapping without on-demand migration cost.
This result demonstrates that \amalgam is a promising cost-efficient solution for LLM acceleration.

\begin{figure}[t]
    \centering
    \includegraphics[width=\columnwidth]{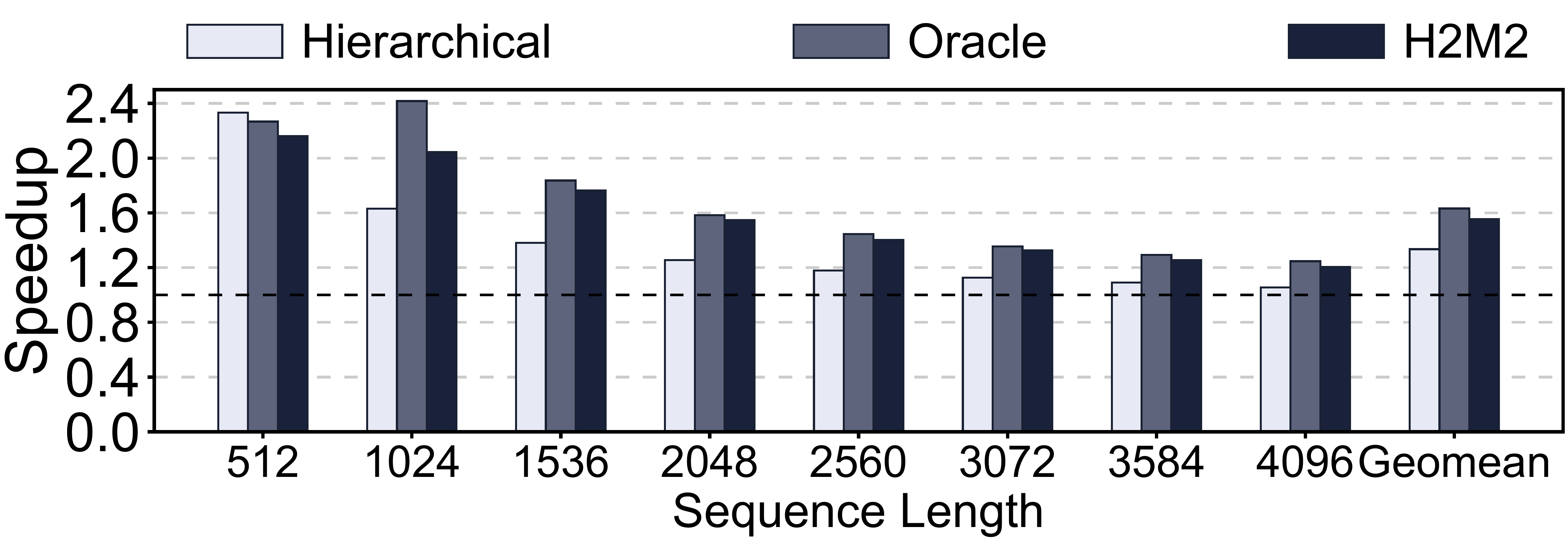}
    \vspace{-4ex}
    \caption{The relative speedup over the baseline (LPDDR-only) for hierarchical memory and asymmetric memory in Chinchilla-70B, batch size 64.}
    \label{fig:eval-hetero-small}
    \vspace{-3ex}
\end{figure}

\noindent
\textbf{Chinchilla-70B:}
Figure~\ref{fig:eval-hetero-small} shows the performance of Chinchilla-70B with an asymmetric memory and alternative memory configurations, represented as a speedup over the baseline.
Note that we use longer sequence lengths and larger batch size for Chinchilla-70B to model a scenario where heterogeneous memory is required due to the lack of bandwidth-centric memory capacity.
Because of the small model size of Chinchilla-70B, both strict hierarchical memory and \amalgam shows higher performance boost over the baseline compared to GPT3-175B.
In addition, strict hierarchical memory outperforms asymmetric memory with the sequence length smaller than 512: When the total footprint of LLM inference task is smaller than the HBM capacity, the performance of strict hierarchical memory become equivalent to the multi-HBM memory without communication cost.
However, this is only a corner case for short sequence length and \amalgam still notably outperforms strict hierarchical memory and closely follows the performance of ideal asymmetric memory in general; strict hierarchical memory shows a 1.33$\times$ speedup over the baseline on average, and \amalgam shows a 1.55$\times$ performance boost, which is 0.95$\times$ of the ideal performance.
This implies that in real-world LLM serving with dynamically growing sequence length, \amalgam is still efficient even with small model size.

\begin{figure}[t]
    \centering
    \includegraphics[width=\columnwidth]{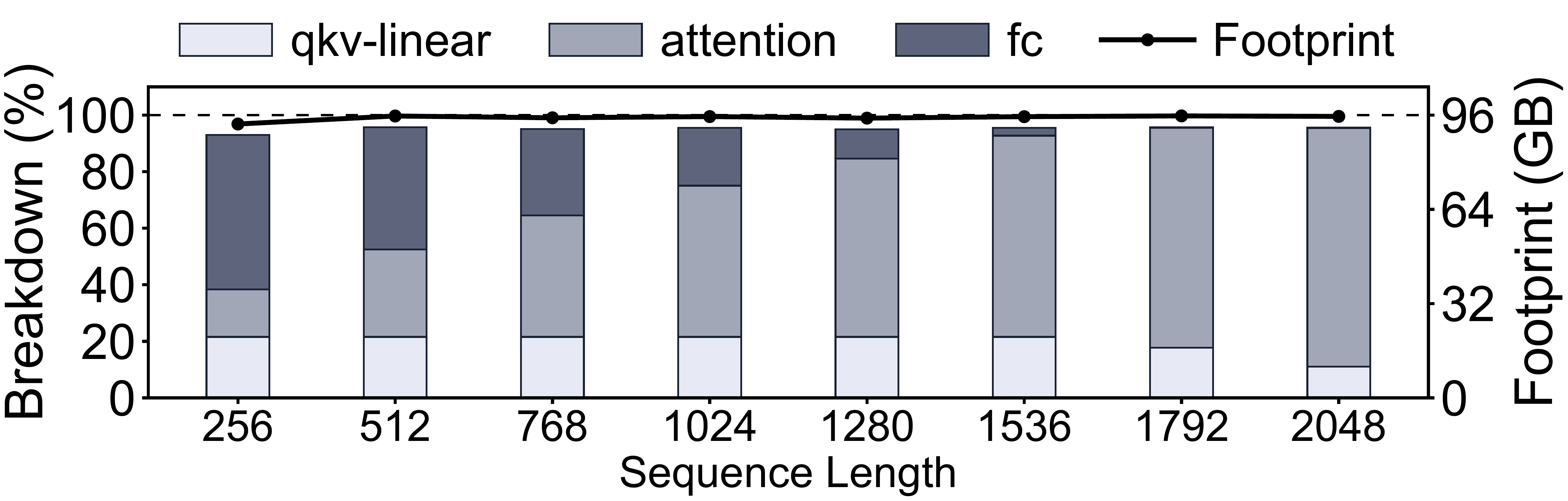}
    \vspace{-4ex}
    \caption{HBM footprint breakdown of \amalgam in GPT3-175B, batch size 32.}
    \label{fig:eval-footprint-breakdown}
    \vspace{-3ex}
\end{figure}

\subsubsection{Analysis with Footprint Breakdown}
Figure~\ref{fig:eval-footprint-breakdown} illustrates the breakdown of HBM utilization during the execution of GPT3-175B with a batch size 32, corresponding to Figure~\ref{fig:eval-hetero-gpt3}.
The bars divided into three colors represent the HBM occupancy for \textit{qkv-linear}, \textit{attention}, and \textit{fc}, while the line plot indicates the overall HBM footprint.
HBM is nearly fully utilized across various sequence lengths as shown in Figure~\ref{fig:eval-footprint-breakdown}, demonstrating the effectiveness of \amalgam's mapping in optimizing HBM utilization. 
Furthermore, \textit{attention} consumes a larger portion of HBM as a sequence length increases, while \textit{fc} exhibits the opposite trend. 
Meanwhile, \textit{qkv-linear} consistently occupies a moderate amount of HBM compared to \textit{attention} and \textit{fc}.
These patterns in HBM utilization demonstrate that \amalgam's mapping strategy effectively adapts to the runtime performance trends discussed in Section~\ref{subsec:tendency}.


\subsubsection{Effect of Grouped-Query Attention (GQA)}
Lllama2-70B employs grouped-query attention (GQA), an effective approach to address memory bandwidth limitations by allowing multiple heads to share the KV cache, with the number of heads determined by the parameter `group size'.
Consequently, GQA exhibits two distinctive characteristics compared to multi-head attention (MHA): (1) the storage and computational overhead for generating K and V tensors is reduced by a factor of 1/(group size), and (2) the size of KV cache in \textit{attention} is also scaled down by the same factor. 
We incorporate the group size parameter into our performance model within \amalgam to account for the impact of GQA.

\begin{figure}[t]
    \centering
    \includegraphics[width=\columnwidth]{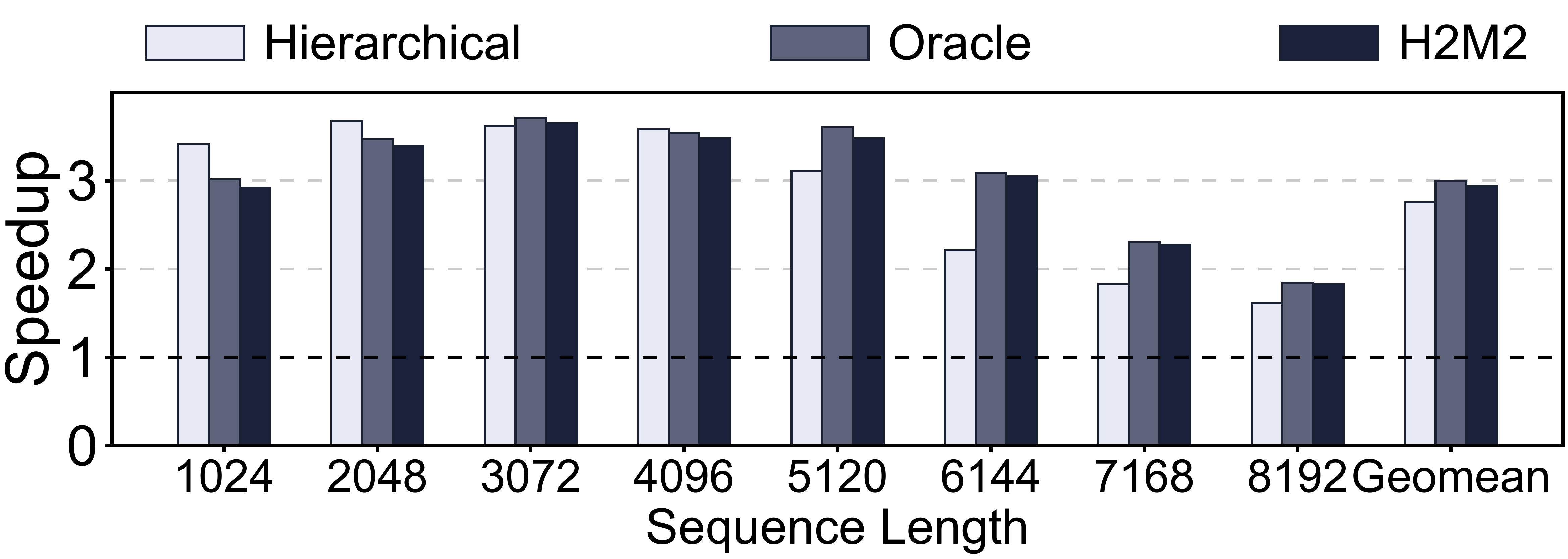}
    \vspace{-4ex}
    \caption{The relative speedup over the baseline (LPDDR-only) for hierarchical memory and asymmetric memory in Llama2-70B, batch size 128.}
    \label{fig:eval-hetero-gqa}
    \vspace{-2ex}
\end{figure}

Figure~\ref{fig:eval-hetero-gqa} illustrates the impact of GQA on utilizing the asymmetric memory architecture.
We use a batch size of 128, and the x-axis and y-axis respectively represent the speedup over the baseline and the sequence length.
Note that a heterogeneous memory architecture becomes meaningful only when the application generates a sufficiently large memory footprint.
Since the memory footprint in GQA is smaller due to the scaled-down size of the KV cache, we use larger batch size and sequence lengths compared to those in Figure~\ref{fig:eval-hetero-small}.
The average speedup over the baseline is 2.75$\times$ for \texttt{Hierarchical}, 3.00$\times$ for \texttt{Oracle}, and 2.94$\times$ for \amalgam, excedding the speedups observed in Chinchilla-70B with a similar parameter size.
This difference can be explained by the effects of GQA in reducing the memory footprint of \textit{attention} and \textit{qkv-linear}, which have the highest priority in HBM usage, resulting in improved HBM utilization.
Although \amalgam generally outperforms \texttt{Hierarchical}, the opposite trend is observed for shorter sequence lengths, as the reduced KV cache size in GQA mitigates the migration cost of \texttt{Hierarchical}.
Although \amalgam requires longer sequence length and larger batch size to be effective in smaller LLMs, this requirement is same as the escape condition for the corner case that the heterogeneous memory system is less effective than single-HBM system.
Overall, the results presented in Figure~\ref{fig:eval-hetero-gqa} show that \amalgam remains effective even when grouped-query attention is applied.


\begin{table}[t]
    \centering
    \footnotesize
    \begin{tabular}{c|ccc}
        \toprule
        \multirow{2}{*}{\textbf{Model}} & \multicolumn{3}{c}{\textbf{Temporal Overhead}} \\
        & \textbf{Abstraction} & \textbf{Mapping} & \textbf{Total} \\
        \midrule
        GPT3-175B & 0.80\% & 2.56\% & 3.36\% \\
        Chinchilla-70B & 1.01\% & 3.76\% & 4.78\% \\
        Llama2-70B & 1.36\% & 0.60\% & 1.96\% \\
        \toprule
    \end{tabular}
    \vspace{-1ex}
    \caption{Average temporal overhead of \amalgam.}
    \label{tab:abstraction-overhead}
    \vspace{-4ex}
\end{table}

\subsubsection{Performance Overheads}
\label{subsubsec:perf_overheads}
While \amalgam follows up the performance of \texttt{Oracle} in all three LLMs used in our evaluation closely, still there exists some gap.
We evaluate the temporal overhead introduced by \textit{memory abstraction} and the reliability of the \textit{greedy mapping policy} to justify the proposed architectural design of \amalgam.
Table~\ref{tab:abstraction-overhead} summarizes the average performance overhead introduced by applying memory abstraction (i.e. adding TLB access and miss latency) and replacing the oracle mapping policy (i.e. always finding the best) to greedy mapping policy.
The performance gap between \amalgam and \texttt{Oracle} in Figure~\ref{fig:eval-hetero-gpt3} and \ref{fig:eval-hetero-small} comes from these performance overhead.
As shown in the first column of the table, the memory abstraction incurs negligible temporal overhead of which is less than 2\% for all three models used in the evaluation.
In addition, the greedy mapping policy adds only 2.56\%, 3.76\%, and 0.60\% additional performance overhead for GPT3-175B, Chinchilla-70B, and Llama2-70B, respectively.
This result shows the robustness of the asymmetric memory architecture with memory abstraction and proposed mapping decision algorithm, with various model sizes and characteristics.

\subsection{Dynamic Sequence Length}
\label{subsec:eval-dynamic}
Experiments of Section~\ref{sec:motiv} and Section~\ref{subsec:speedup} assume all requests in a batch have equal sequence length without early termination of any requests.
Here, to evaluate the performance of \amalgam and the efficiency of greedy mapping policy in realistic and dynamic situation, we consider another scenario where requests in a batch can have different sequence length and can be terminated in random moment.
We assume that requests are consecutively fed to the system, and we fix the batch size to 32.
For every iteration, each request can either generate the next token (i.e. sequence length increases by 1), or can be finished and replaced to the new request (i.e. sequence length changes randomly).
Such a dynamic scenario leads KV cache size to fluctuate in runtime, making optimal mapping change frequently.

\begin{figure}[t]
    \centering
    \includegraphics[width=\columnwidth]{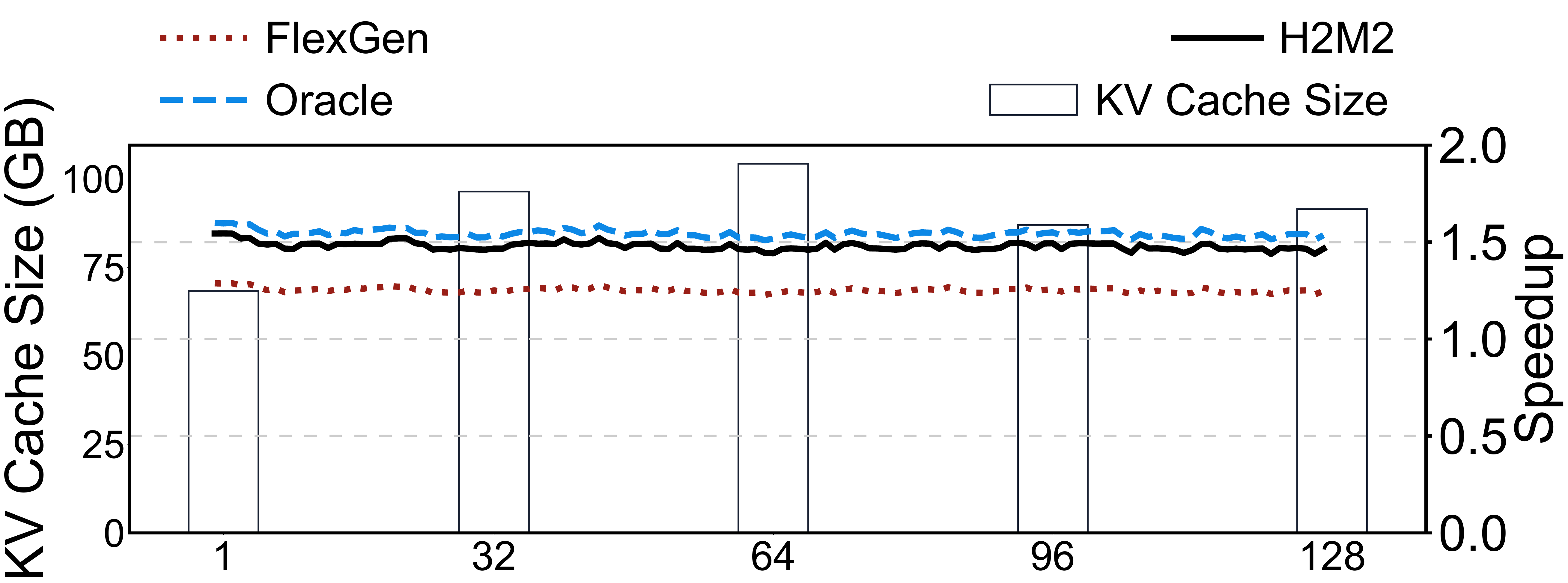}
    \vspace{-4ex}
    \caption{The relative speedup over the baseline for asymmetric memory, in dynamic sequence length scenario in GPT3-175B.}
    \label{fig:eval-dynamic}
    \vspace{-2ex}
\end{figure}

Figure~\ref{fig:eval-dynamic} visualizes the performance of \amalgam, \texttt{Oracle}, and \texttt{FlexGen} with dynamic sequence length scenario, measured during 128 iterations in GPT3-175B.
The x-axis is the number of cumulative iterations, and the y-axis is a speedup over the baseline.
Moreover, we measure the total size of KV cache for five checkpoints: the 1st, 32nd, 64th, 96th, and 128th iteration, respectively.
As shown in Figure~\ref{fig:eval-dynamic}, \amalgam shows a stable speedup over the baseline even with the dynamic change of KV cache size as well as following the performance of \texttt{Oracle} closely
- While \texttt{FlexGen} only reports 1.25$\times$ speedup over the baseline on average, \texttt{H2D2} shows 1.48$\times$ speedup over the baseline on average, reaching 0.96$\times$ performance of \texttt{Oracle}.
This implies the robustness of \textit{greedy mapping policy}.
Note that \texttt{FlexGen} reports larger performance gap under \amalgam in Figure~\ref{fig:eval-dynamic} compared to Figure~\ref{fig:hetero-mapping-vs-flexgen}, due to the difference of the batch size.

\subsection{Sensitivity Study}
\label{subsec:sensitivity}
To demonstrate the robustness of \abcmem design to different hardware configurations, we conduct a sensitivity study with eight different hardware configurations outlined in Table~\ref{tab:sensitivity_config}.
While \texttt{Original} is an original memory configuration of Table~\ref{tab:memconfig}, the other  configurations are created by changing one of the configuration parameters from \texttt{Original}: Either deducting (\texttt{Less}) or increasing (\texttt{More}) HBM capacity, HBM bandwidth, LPDDR bandwidth, HBM-side computational power, and LPDDR-side computational power.

\begin{table}[t]
    \centering
    \footnotesize
    \begin{tabular}{c|cc}
        \toprule
        \textbf{Name} & \textbf{Component} & \textbf{Configuration} \\ 
        \midrule
        \texttt{Original} & \multicolumn{2}{c}{No modification} \\
        \texttt{HBMcap-Less} & HBM capacity & 48GB (0.5$\times$) \\
        \texttt{HBMcap-More} & HBM capacity & 192GB (2$\times$) \\
        \texttt{HBMbw-Less} & HBM bandwidth & 2.25TB/s (0.75$\times$) \\
        \texttt{HBMbw-More} & HBM bandwidth & 4TB/s (1.3$\times$) \\
        \texttt{LPDDRbw-Less} & LPDDR bandwidth & 408GB/s (0.75$\times$) \\
        \texttt{LPDDRbw-More} & LPDDR bandwidth & 680GB/s (1.25$\times$) \\
        \texttt{HBMChip-More} & HBM-side compute units & 2 chips (2$\times$) \\
        \texttt{LPDDRChip-More} & LPDDR-side compute units & 2 chips (2$\times$) \\
        \toprule
    \end{tabular}
    \caption{Memory configuration variants for sensitivity study.}
    \label{tab:sensitivity_config}
    \vspace{-6ex}
\end{table}

\begin{figure}[t]
    \centering
    \includegraphics[width=\columnwidth]{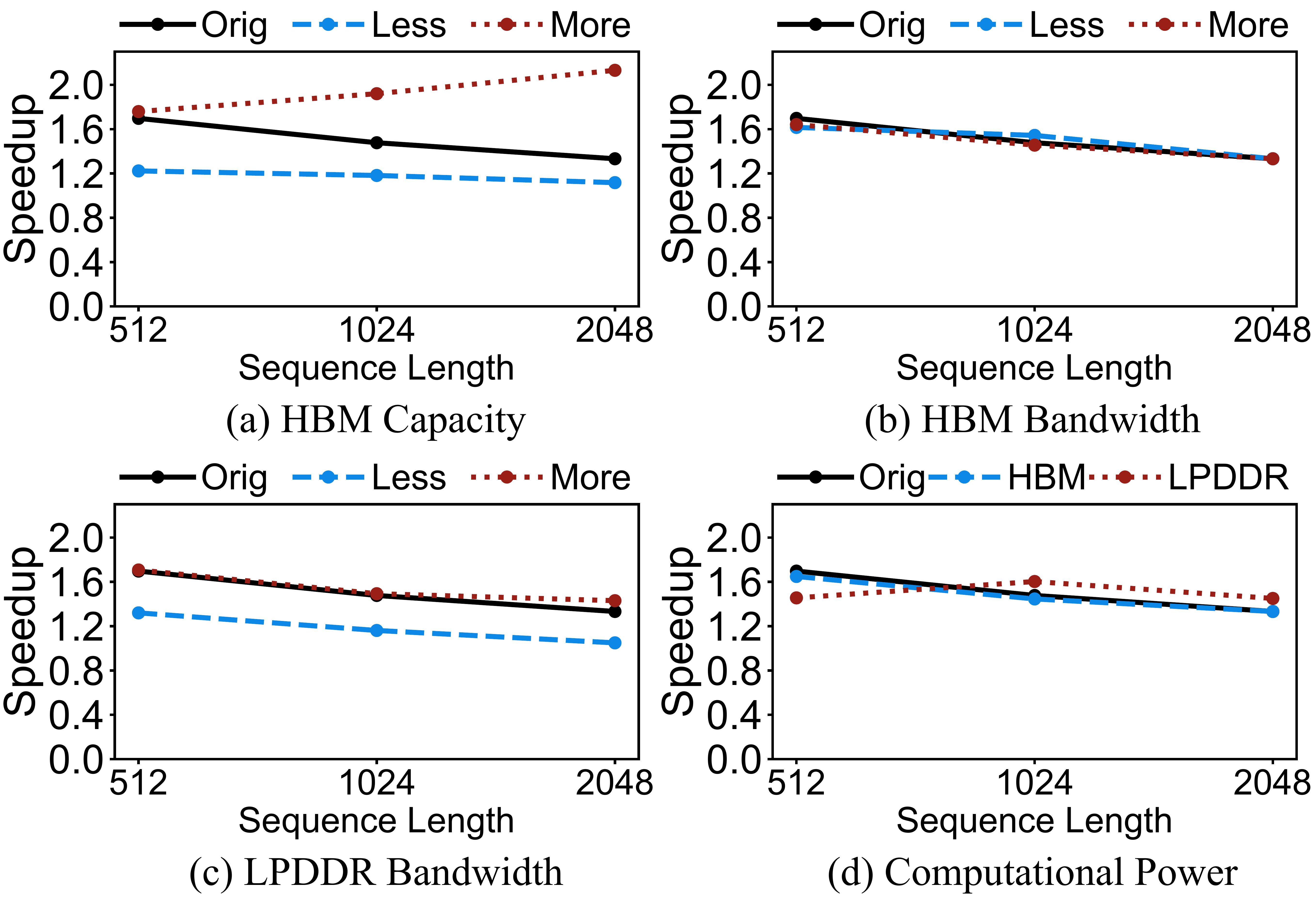}
     \vspace{-0.3in}
    \caption{Sensitivity study for \amalgam with GPT3-175B.}
    \label{fig:eval-sensitivity}
    \vspace{-2ex}
\end{figure}

Figure~\ref{fig:eval-sensitivity} presents the performance trend of various hardware configurations compared to \texttt{Original}.
As shown in Figure~\ref{fig:eval-sensitivity}(a), the relative performance of \amalgam over the baseline is highly sensitive to the capacity of HBM (\texttt{HBMcap}).
This is because the capacity of HBM is directly related to the amount of kernels that can be mapped to HBM, deciding the performance of overall system.
The bandwidth of LPDDR (\texttt{LPDDRbw}) and the computational power of LPDDR-side accelerator (\texttt{LPDDRChip}) also shows some impacts on performance, as illustrated in Figure~\ref{fig:eval-sensitivity}(c) and (d).
As the performance of \amalgam is bounded to the LPDDR-side execution, the change of LPDDR-side configuration affects the performance of overall system. 
Meanwhile, the sensitivity of performance to other parameters such as HBM bandwidth are almost negligible, as shown in  Figure~\ref{fig:eval-sensitivity}(b).
This is because other parameters do not make significant impact on the performance of the critical path.

Despite the variance on performance trends of eight alternative configurations, analyzing the distinctions in each configuration indicates the reasonableness of diverse patterns.
This sensitivity study highlights the reliability of deploying the \abcmem for efficient LLM acceleration in various hardware configurations.

\subsection{Comparison with Multi-HBM System}
\label{subsec:vs-multi-hbm}
To compare \amalgam with multi-HBM system, we introduce a new configuration called \texttt{8-HBM}, an HBM-only system with eight HBM devices that offers total capacity of 768GB.
As multiple HBM devices require communication between memory modules, we add communication cost to \texttt{8-HBM} which was measured by profiling multi-GPU system with eight NVIDIA A100 GPUs.

\begin{figure}[t]
    \centering
    \includegraphics[width=\columnwidth]{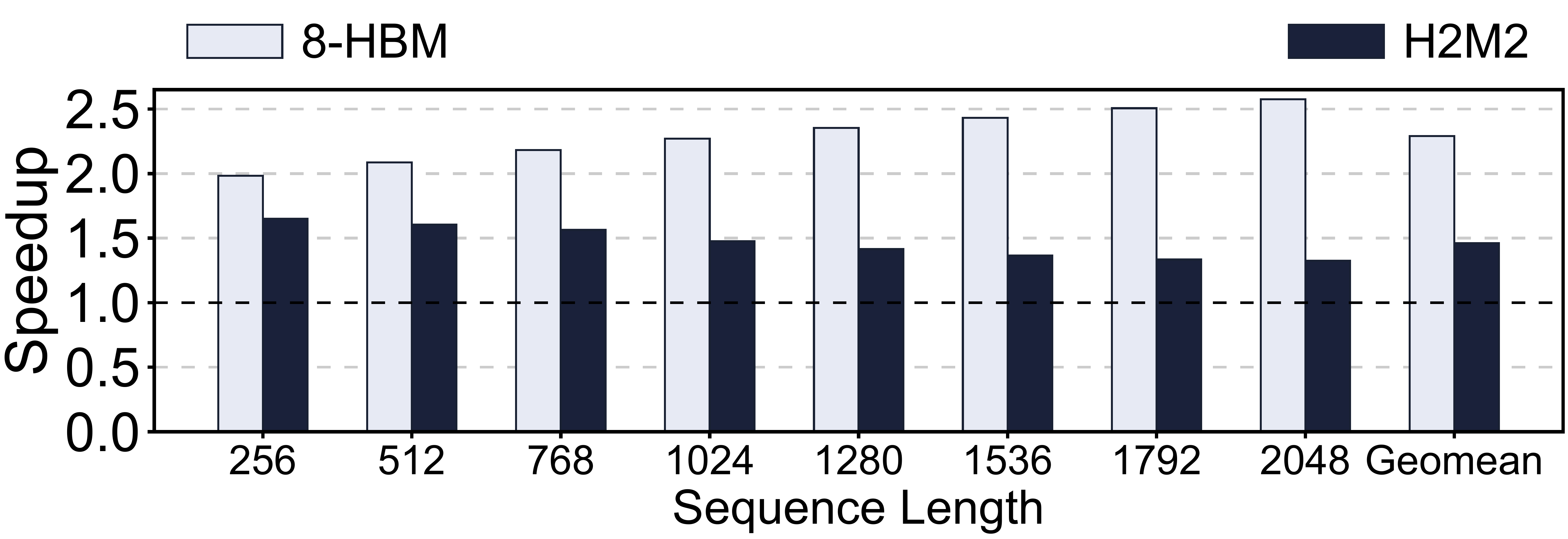}
    \vspace{-5ex}
    \caption{The performance improvement of \amalgam and multi-HBM system (\texttt{8-HBM}) over the baseline (LPDDR-only) for GPT3-175B, batch size 32.}
    \label{fig:eval-vs-hbm}
    \vspace{-2ex}
\end{figure}

\begin{figure}[t]
    \centering
    \includegraphics[width=\columnwidth]{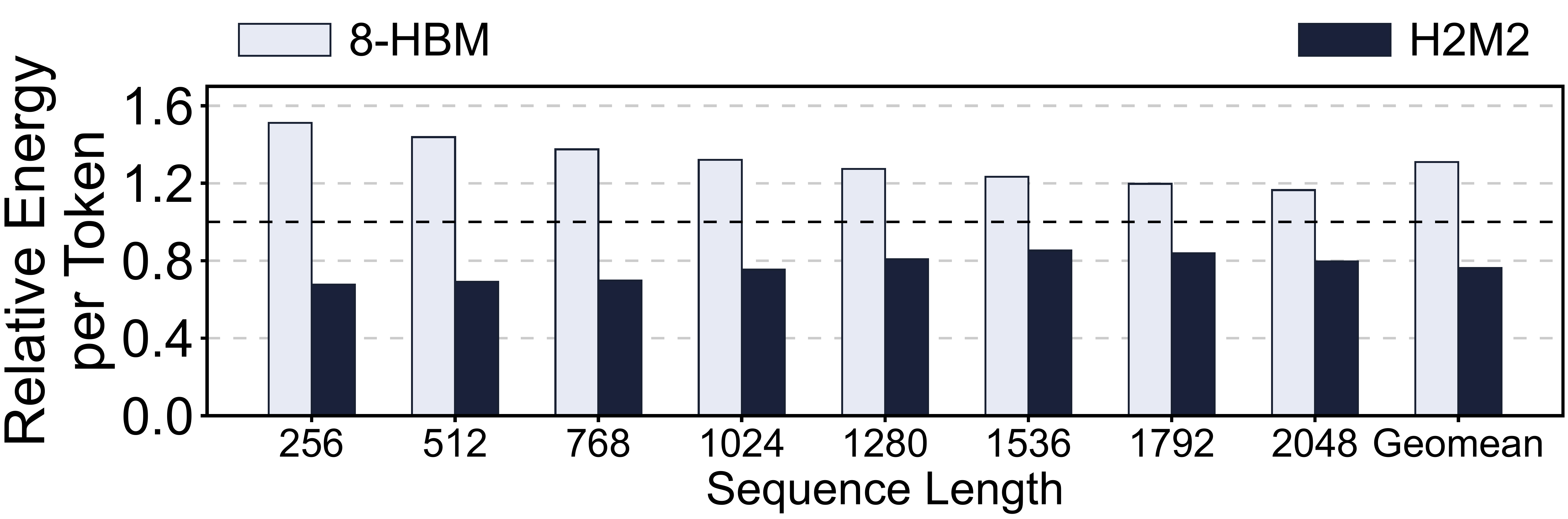}
    \vspace{-5ex}
    \caption{The relative memory energy per token \amalgam and \texttt{8-HBM} over the baseline (LPDDR-only) for GPT3-175B, batch size 32. Lower is better.}
    \label{fig:eval-energy-per-tok}
    \vspace{-3ex}
\end{figure}

\noindent
\textbf{Speedup:}
Figure~\ref{fig:eval-vs-hbm} compares the relative speedup of \amalgam and \texttt{8-HBM} over the baseline LPDDR-only system for GPT3-175B, with batch size of 32.
While \amalgam achieves average speedup of 1.46$\times$ over the baseline, \texttt{8-HBM} outperforms the baseline with an average speedup of 2.29$\times$, which can be trasnlated into 1.57$\times$ additional speedup over \amalgam.
Such performance gap between \texttt{8-HBM} and \amalgam comes from from the limited capacity of HBM as well as low locality of LLM.

\noindent
\textbf{Energy Efficiency:}
To evaluate the energy efficiency, we compare the energy consumption per token of \amalgam with that of \texttt{8-HBM}.
Note that the computational capability of both \amalgam and \texttt{8-HBM} remains unchanged, and memory access overhead generally dominates power consumption compared to computational overhead~\cite{maestro, accelergy, eyeriss, tetris}.
Accordingly, our static power model focuses solely on memory accesses, assuming equal energy consumptions for read and write operations.
The relative energy consumption of HBM and LPDDR in our static power model is derived from the prior work~\cite{cxlpnm}.
Figure~\ref{fig:eval-energy-per-tok} visualizes the relative memory energy consumption per token of \amalgam and \texttt{8-HBM} over the baseline LPDDR-only system.
As shown in the figure, \amalgam's energy consumption per token is 0.76$\times$ of the baseline on average, while \texttt{8-HBM} reports 1.31$\times$ of the baseline on average.
Considering that the relative energy consumption of the baseline LPDDR-only system is translated to 1, this result justifies the asymmetric memory system design for LLMs as \amalgam is more energy-efficient than both multi-HBM and LPDDR-only systems.
%

\noindent
\textbf{Cost and Scalability:}
Recall Section~\ref{subsec:background-hm-llm} , larger memory capacity requires more devices with multi-HBM systems, leading multi-HBM systems to become increasingly affected by communication overhead.
Because of this, the cost-to-performance efficiency of multi-HBM systems tends to degrade as the system's memory capacity increases~\cite{alpa}.
In contrast, asymmetric memory systems combine highly scalable LPDDR with less scalable HBM, allowing memory capacity to be expanded on the LPDDR side.
Based on this design, \amalgam makes it easier to mitigate the decline in cost-to-performance efficiency as memory capacity increases.
\section{Related Work}

\noindent
\textbf{LLM Serving Systems: }
The research into systems for efficient LLM execution has been primarily carried out on GPUs.
Orca improved the throughput of LLM inference by adopting iteration-level scheduling \cite{orca}.
FlashLLM reduced data traffic by using flash memory and accessing data within larger granularity~\cite{flashllm}.
FlashAttention lowered memory usage and I/O costs of attention layers through input block organization and kernel optimization~\cite{flashattention}.
PowerInfer alleviates GPU memory limitations by uploading only the active (hot) neurons of the model’s weights to the GPU, while delegating the remaining computations to the host CPU~\cite{powerinfer}.
ALISA and InfiniGen reduces memory demands and swap overhead in LLM inference by employing sparse attention, processing only critical key-value pairs during attention operations~\cite{alisa, infinigen}.

\noindent
\textbf{LLM Accelerators: }
LLM accelerators have been deeply researched for power efficiency and performance via specialized hardware and parallel architectures.
DFX implemented a specialized hardware accelerator architecture for LLM with FPGA~\cite{dfx}.
FlightLLM used mixed precision to place activations in the on-chip memory during the decode stage to accelerate LLM inference~\cite{flightllm}.

\noindent
\textbf{Near-memory Processing for LLMs: }
Several studies have investigated near-memory processing to accelerate LLMs.
IANUS, NeuPIMs and AttAcc integrated GEMV-optimized PIM technology to accelerate the attention layer efficiently~\cite{ianus, neupims, attacc}.
Smart-Infinity utilized computational storage devices for updating parameters in LLM training~\cite{smartinfinity}.

\section{Conclusion}
With asymmetric memory for LLM acceleration, this study investigated the mapping decision algorithm, and proposed a memory abstraction scheme for the efficient management.
The propose \abcmem outperformed the conventional LPDDR-based homogeneous memory system by 1.46$\times$ speedup in GPT3-175B on average, and traditional strict hierarchical memory by 1.36$\times$.


\balance
\bibliographystyle{ACM-Reference-Format} 
\bibliography{references}


\begin{thebibliography}{54}


\ifx \showCODEN    \undefined \def \showCODEN     #1{\unskip}     \fi
\ifx \showDOI      \undefined \def \showDOI       #1{#1}\fi
\ifx \showISBNx    \undefined \def \showISBNx     #1{\unskip}     \fi
\ifx \showISBNxiii \undefined \def \showISBNxiii  #1{\unskip}     \fi
\ifx \showISSN     \undefined \def \showISSN      #1{\unskip}     \fi
\ifx \showLCCN     \undefined \def \showLCCN      #1{\unskip}     \fi
\ifx \shownote     \undefined \def \shownote      #1{#1}          \fi
\ifx \showarticletitle \undefined \def \showarticletitle #1{#1}   \fi
\ifx \showURL      \undefined \def \showURL       {\relax}        \fi
\providecommand\bibfield[2]{#2}
\providecommand\bibinfo[2]{#2}
\providecommand\natexlab[1]{#1}
\providecommand\showeprint[2][]{arXiv:#2}

\bibitem[Alizadeh et~al\mbox{.}(2023)]%
        {flashllm}
\bibfield{author}{\bibinfo{person}{Keivan Alizadeh}, \bibinfo{person}{Iman
  Mirzadeh}, \bibinfo{person}{Dmitry Belenko}, \bibinfo{person}{Karen
  Khatamifard}, \bibinfo{person}{Minsik Cho}, \bibinfo{person}{Carlo~C
  Del~Mundo}, \bibinfo{person}{Mohammad Rastegari}, {and}
  \bibinfo{person}{Mehrdad Farajtabar}.} \bibinfo{year}{2023}\natexlab{}.
\newblock \showarticletitle{Llm in a flash: Efficient large language model
  inference with limited memory}.
\newblock \bibinfo{journal}{\emph{arXiv preprint arXiv:2312.11514}}
  (\bibinfo{year}{2023}).
\newblock


\bibitem[Aminabadi et~al\mbox{.}(2022)]%
        {deepspeed}
\bibfield{author}{\bibinfo{person}{Reza~Yazdani Aminabadi},
  \bibinfo{person}{Samyam Rajbhandari}, \bibinfo{person}{Ammar~Ahmad Awan},
  \bibinfo{person}{Cheng Li}, \bibinfo{person}{Du Li}, \bibinfo{person}{Elton
  Zheng}, \bibinfo{person}{Olatunji Ruwase}, \bibinfo{person}{Shaden Smith},
  \bibinfo{person}{Minjia Zhang}, \bibinfo{person}{Jeff Rasley},
  {et~al\mbox{.}}} \bibinfo{year}{2022}\natexlab{}.
\newblock \showarticletitle{Deepspeed-inference: enabling efficient inference
  of transformer models at unprecedented scale}. In
  \bibinfo{booktitle}{\emph{SC22: International Conference for High Performance
  Computing, Networking, Storage and Analysis}}. IEEE, \bibinfo{pages}{1--15}.
\newblock


\bibitem[Anil et~al\mbox{.}(2023)]%
        {palm2}
\bibfield{author}{\bibinfo{person}{Rohan Anil}, \bibinfo{person}{Andrew~M.
  Dai}, \bibinfo{person}{Orhan Firat}, \bibinfo{person}{Melvin Johnson},
  \bibinfo{person}{Dmitry Lepikhin}, \bibinfo{person}{Alexandre Passos},
  \bibinfo{person}{Siamak Shakeri}, \bibinfo{person}{Emanuel Taropa},
  \bibinfo{person}{Paige Bailey}, \bibinfo{person}{Zhifeng Chen},
  {et~al\mbox{.}}} \bibinfo{year}{2023}\natexlab{}.
\newblock \showarticletitle{{PaLM 2 Technical Report}}.
\newblock \bibinfo{journal}{\emph{arXiv preprint arXiv:2305.10403}}
  (\bibinfo{year}{2023}).
\newblock
\showeprint[arxiv]{2305.10403}~[cs.CL]


\bibitem[Borzunov et~al\mbox{.}(2024)]%
        {petals}
\bibfield{author}{\bibinfo{person}{Alexander Borzunov}, \bibinfo{person}{Max
  Ryabinin}, \bibinfo{person}{Artem Chumachenko}, \bibinfo{person}{Dmitry
  Baranchuk}, \bibinfo{person}{Tim Dettmers}, \bibinfo{person}{Younes Belkada},
  \bibinfo{person}{Pavel Samygin}, {and} \bibinfo{person}{Colin~A Raffel}.}
  \bibinfo{year}{2024}\natexlab{}.
\newblock \showarticletitle{Distributed Inference and Fine-tuning of Large
  Language Models Over The Internet}.
\newblock \bibinfo{journal}{\emph{Advances in Neural Information Processing
  Systems}}  \bibinfo{volume}{36} (\bibinfo{year}{2024}).
\newblock


\bibitem[Brown et~al\mbox{.}(2020)]%
        {gpt3}
\bibfield{author}{\bibinfo{person}{Tom~B. Brown}, \bibinfo{person}{Benjamin
  Mann}, \bibinfo{person}{Nick Ryder}, \bibinfo{person}{Melanie Subbiah},
  \bibinfo{person}{Jared Kaplan}, \bibinfo{person}{Prafulla Dhariwal},
  \bibinfo{person}{Arvind Neelakantan}, \bibinfo{person}{Pranav Shyam},
  \bibinfo{person}{Girish Sastry}, \bibinfo{person}{Amanda Askell},
  \bibinfo{person}{Sandhini Agarwal}, \bibinfo{person}{Ariel Herbert-Voss},
  \bibinfo{person}{Grechen Krueger}, \bibinfo{person}{Tom Henighan},
  \bibinfo{person}{Rewon Child}, \bibinfo{person}{Aditya Ramesh},
  \bibinfo{person}{Daniel~M. Ziegler}, \bibinfo{person}{Jeffrey Wu},
  \bibinfo{person}{Clemens Winter}, \bibinfo{person}{Christopher Hesse},
  \bibinfo{person}{Mark Chen}, \bibinfo{person}{Eric Sigler},
  \bibinfo{person}{Mateusz Litwin}, \bibinfo{person}{Scott Gray},
  \bibinfo{person}{Benjamin Chess}, \bibinfo{person}{Jack Clark},
  \bibinfo{person}{Christopher Berner}, \bibinfo{person}{Sam McCandlish},
  \bibinfo{person}{Alec Radford}, \bibinfo{person}{Ilya Sutskever}, {and}
  \bibinfo{person}{Dario Amodei}.} \bibinfo{year}{2020}\natexlab{}.
\newblock \showarticletitle{{Language Models are Few-Shot Learners}}.
\newblock \bibinfo{journal}{\emph{Advances in neural information processing
  systems}}  \bibinfo{volume}{33} (\bibinfo{year}{2020}),
  \bibinfo{pages}{1877--1901}.
\newblock


\bibitem[Chen et~al\mbox{.}(2016)]%
        {eyeriss}
\bibfield{author}{\bibinfo{person}{Yu-Hsin Chen}, \bibinfo{person}{Joel Emer},
  {and} \bibinfo{person}{Vivienne Sze}.} \bibinfo{year}{2016}\natexlab{}.
\newblock \showarticletitle{Eyeriss: A spatial architecture for
  energy-efficient dataflow for convolutional neural networks}.
\newblock \bibinfo{journal}{\emph{ACM SIGARCH computer architecture news}}
  \bibinfo{volume}{44}, \bibinfo{number}{3} (\bibinfo{year}{2016}),
  \bibinfo{pages}{367--379}.
\newblock


\bibitem[Chowdhery et~al\mbox{.}(2022)]%
        {palm}
\bibfield{author}{\bibinfo{person}{Aakanksha Chowdhery},
  \bibinfo{person}{Sharan Narang}, \bibinfo{person}{Jacob Devlin},
  \bibinfo{person}{Maarten Bosma}, \bibinfo{person}{Gaurav Mishra},
  \bibinfo{person}{Adam Roberts}, \bibinfo{person}{Paul Barham},
  \bibinfo{person}{Hyung~Won Chung}, \bibinfo{person}{Charles Sutton},
  \bibinfo{person}{Sebastian Gehrmann}, {et~al\mbox{.}}}
  \bibinfo{year}{2022}\natexlab{}.
\newblock \showarticletitle{Palm: Scaling language modeling with pathways}.
\newblock \bibinfo{journal}{\emph{arXiv preprint arXiv:2204.02311}}
  (\bibinfo{year}{2022}).
\newblock


\bibitem[Dao et~al\mbox{.}(2022)]%
        {flashattention}
\bibfield{author}{\bibinfo{person}{Tri Dao}, \bibinfo{person}{Dan Fu},
  \bibinfo{person}{Stefano Ermon}, \bibinfo{person}{Atri Rudra}, {and}
  \bibinfo{person}{Christopher R{\'e}}.} \bibinfo{year}{2022}\natexlab{}.
\newblock \showarticletitle{Flashattention: Fast and memory-efficient exact
  attention with io-awareness}.
\newblock \bibinfo{journal}{\emph{Advances in Neural Information Processing
  Systems}}  \bibinfo{volume}{35} (\bibinfo{year}{2022}),
  \bibinfo{pages}{16344--16359}.
\newblock


\bibitem[Gao et~al\mbox{.}(2017)]%
        {tetris}
\bibfield{author}{\bibinfo{person}{Mingyu Gao}, \bibinfo{person}{Jing Pu},
  \bibinfo{person}{Xuan Yang}, \bibinfo{person}{Mark Horowitz}, {and}
  \bibinfo{person}{Christos Kozyrakis}.} \bibinfo{year}{2017}\natexlab{}.
\newblock \showarticletitle{Tetris: Scalable and efficient neural network
  acceleration with 3d memory}. In \bibinfo{booktitle}{\emph{Proceedings of the
  Twenty-Second International Conference on Architectural Support for
  Programming Languages and Operating Systems}}. \bibinfo{pages}{751--764}.
\newblock


\bibitem[Ghose et~al\mbox{.}(2019)]%
        {dram-analysis}
\bibfield{author}{\bibinfo{person}{Saugata Ghose}, \bibinfo{person}{Tianshi
  Li}, \bibinfo{person}{Nastaran Hajinazar}, \bibinfo{person}{Damla~Senol
  Cali}, {and} \bibinfo{person}{Onur Mutlu}.} \bibinfo{year}{2019}\natexlab{}.
\newblock \showarticletitle{{Demystifying Complex Workload-DRAM Interactions:
  An Experimental Study}}.
\newblock  \bibinfo{volume}{3}, \bibinfo{number}{3} (\bibinfo{year}{2019}).
\newblock


\bibitem[Google(2018)]%
        {cloudtpu}
\bibfield{author}{\bibinfo{person}{Google}.} \bibinfo{year}{2018}\natexlab{}.
\newblock \bibinfo{title}{Cloud{TPU}}.
\newblock
  \bibinfo{howpublished}{https://cloud.google.com/tpu/docs/system-architecture-tpu-vm}.
\newblock


\bibitem[Gu and Dao(2023)]%
        {mamba}
\bibfield{author}{\bibinfo{person}{Albert Gu} {and} \bibinfo{person}{Tri Dao}.}
  \bibinfo{year}{2023}\natexlab{}.
\newblock \showarticletitle{Mamba: Linear-time sequence modeling with selective
  state spaces}.
\newblock \bibinfo{journal}{\emph{arXiv preprint arXiv:2312.00752}}
  (\bibinfo{year}{2023}).
\newblock


\bibitem[Heo et~al\mbox{.}(2024)]%
        {neupims}
\bibfield{author}{\bibinfo{person}{Guseul Heo}, \bibinfo{person}{Sangyeop Lee},
  \bibinfo{person}{Jaehong Cho}, \bibinfo{person}{Hyunmin Choi},
  \bibinfo{person}{Sanghyeon Lee}, \bibinfo{person}{Hyungkyu Ham},
  \bibinfo{person}{Gwangsun Kim}, \bibinfo{person}{Divya Mahajan}, {and}
  \bibinfo{person}{Jongse Park}.} \bibinfo{year}{2024}\natexlab{}.
\newblock \showarticletitle{{NeuPIMs: A NPU-PIM Heterogeneous Acceleration for
  Batched Inference of Large Language Model}}. In
  \bibinfo{booktitle}{\emph{Proceedings of the 29th ACM International
  Conference on Architectural Support for Programming Languages and Operating
  Systems (ASPLOS'24)}}.
\newblock


\bibitem[Hoffmann et~al\mbox{.}(2022)]%
        {chinchilla}
\bibfield{author}{\bibinfo{person}{Jordan Hoffmann}, \bibinfo{person}{Sebastian
  Borgeaud}, \bibinfo{person}{Arthur Mensch}, \bibinfo{person}{Elena
  Buchatskaya}, \bibinfo{person}{Trevor Cai}, \bibinfo{person}{Eliza
  Rutherford}, \bibinfo{person}{Diego de~Las Casas}, \bibinfo{person}{Lisa~Anne
  Hendricks}, \bibinfo{person}{Johannes Welbi}, \bibinfo{person}{Aidan Clark},
  \bibinfo{person}{Tom Hennigan}, \bibinfo{person}{Eric Noland},
  \bibinfo{person}{Katie Millican}, \bibinfo{person}{George van~den Driessche},
  \bibinfo{person}{Bogdan Damoc}, \bibinfo{person}{Aurelia Guy},
  \bibinfo{person}{Simon Osindero}, \bibinfo{person}{Karen Simonyan},
  \bibinfo{person}{Erich Elsen}, \bibinfo{person}{Jaek~W. Rae},
  \bibinfo{person}{Oriol Vinyals}, {and} \bibinfo{person}{Laurent Sifre}.}
  \bibinfo{year}{2022}\natexlab{}.
\newblock \showarticletitle{{Training Compute-Optimal Large Language Models}}.
\newblock \bibinfo{journal}{\emph{arXiv preprint arXiv:2203.15556}}
  (\bibinfo{year}{2022}).
\newblock


\bibitem[Hong et~al\mbox{.}(2022)]%
        {dfx}
\bibfield{author}{\bibinfo{person}{Seongmin Hong}, \bibinfo{person}{Seungjae
  Moon}, \bibinfo{person}{Junsoo Kim}, \bibinfo{person}{Sungjae Lee},
  \bibinfo{person}{Minsub Kim}, \bibinfo{person}{Dongsoo Lee}, {and}
  \bibinfo{person}{Joo-Young Kim}.} \bibinfo{year}{2022}\natexlab{}.
\newblock \showarticletitle{{DFX: A Low-latency Multi-FPGA Applicance for
  Accelerating Transformer-based Text Generation}}. In
  \bibinfo{booktitle}{\emph{2022 55th Annual IEEE/ACM International Symposium
  on Microarchitecture (MICRO)}}.
\newblock


\bibitem[Hwang et~al\mbox{.}(2023)]%
        {mnpusim}
\bibfield{author}{\bibinfo{person}{Soojin Hwang}, \bibinfo{person}{Sunho Lee},
  \bibinfo{person}{Jungwoo Kim}, \bibinfo{person}{Hongbeen Kim}, {and}
  \bibinfo{person}{Jaehyuk Huh}.} \bibinfo{year}{2023}\natexlab{}.
\newblock \showarticletitle{{mNPUsim: Evaluating the Effect of Sharing
  Resources with Multi-core NPUs}}. In \bibinfo{booktitle}{\emph{2023 IEEE
  International Symposium on Workload Charcterization (IISWC)}}.
\newblock


\bibitem[Hyun et~al\mbox{.}(2020)]%
        {neummu}
\bibfield{author}{\bibinfo{person}{Bongjoon Hyun}, \bibinfo{person}{Youngeun
  Kwon}, \bibinfo{person}{Yujeong Choi}, \bibinfo{person}{John Kim}, {and}
  \bibinfo{person}{Minsoo Rhu}.} \bibinfo{year}{2020}\natexlab{}.
\newblock \showarticletitle{{NeuMMU: Architectural Support for Efficient
  Address Translations in Neural Processing Units}}. In
  \bibinfo{booktitle}{\emph{Proceedings of the 25th ACM International
  Conference on Architectural Support for Programming Languages and Operating
  Systems (ASPLOS'20)}}. \bibinfo{pages}{1109--1124}.
\newblock


\bibitem[Jang et~al\mbox{.}(2024)]%
        {smartinfinity}
\bibfield{author}{\bibinfo{person}{Hongsun Jang}, \bibinfo{person}{Jaeyong
  Song}, \bibinfo{person}{Jaewon Jung}, \bibinfo{person}{Jaeyoung Park},
  \bibinfo{person}{Youngsok Kim}, {and} \bibinfo{person}{Jinho Lee}.}
  \bibinfo{year}{2024}\natexlab{}.
\newblock \showarticletitle{Smart-Infinity: Fast Large Language Model Training
  using Near-Storage Processing on a Real System}. In
  \bibinfo{booktitle}{\emph{2024 IEEE International Symposium on
  High-Performance Computer Architecture (HPCA)}}. IEEE,
  \bibinfo{pages}{345--360}.
\newblock


\bibitem[Jeong et~al\mbox{.}(2023)]%
        {deepplan}
\bibfield{author}{\bibinfo{person}{Jinwoo Jeong}, \bibinfo{person}{Seungsu
  Baek}, {and} \bibinfo{person}{Jeongseob Ahn}.}
  \bibinfo{year}{2023}\natexlab{}.
\newblock \showarticletitle{{Fast and Efficient Model Serving Using Multi-GPUs
  with Direct-Host-Access}}. In \bibinfo{booktitle}{\emph{Proceedings of the
  Eighteenth European Conference on Computer Systems}} (Rome, Italy)
  \emph{(\bibinfo{series}{EuroSys '23})}. \bibinfo{pages}{249–265}.
\newblock
\urldef\tempurl%
\url{https://doi.org/10.1145/3552326.3567508}
\showDOI{\tempurl}


\bibitem[Jun et~al\mbox{.}(2017)]%
        {hbm-arch}
\bibfield{author}{\bibinfo{person}{Hongshin Jun}, \bibinfo{person}{Jinhee Cho},
  \bibinfo{person}{Kangseol Lee}, \bibinfo{person}{Ho-Young Son},
  \bibinfo{person}{Kwiwook Kim}, \bibinfo{person}{Hanho Jin}, {and}
  \bibinfo{person}{Keith Kim}.} \bibinfo{year}{2017}\natexlab{}.
\newblock \showarticletitle{{HBM (High Bandwidth Memory) DRAM Technology and
  Architecture}}. In \bibinfo{booktitle}{\emph{2017 IEEE International Memory
  Workshop}}.
\newblock


\bibitem[Kao et~al\mbox{.}(2023)]%
        {flat}
\bibfield{author}{\bibinfo{person}{Sheng-Chun Kao}, \bibinfo{person}{Suvinay
  Subramanian}, \bibinfo{person}{Gaurav Agrawal}, \bibinfo{person}{Amir
  Yazdanbakhsh}, {and} \bibinfo{person}{Tushar Krishna}.}
  \bibinfo{year}{2023}\natexlab{}.
\newblock \showarticletitle{FLAT: An Optimized Dataflow for Mitigating
  Attention Bottlenecks}. In \bibinfo{booktitle}{\emph{Proceedings of the 28th
  ACM International Conference on Architectural Support for Programming
  Languages and Operating Systems, Volume 2}}. \bibinfo{pages}{295--310}.
\newblock


\bibitem[Korthikanti et~al\mbox{.}(2023)]%
        {seqparallelism2}
\bibfield{author}{\bibinfo{person}{Vijay~Anand Korthikanti},
  \bibinfo{person}{Jared Casper}, \bibinfo{person}{Sangkug Lym},
  \bibinfo{person}{Lawrence McAfee}, \bibinfo{person}{Michael Andersch},
  \bibinfo{person}{Mohammad Shoeybi}, {and} \bibinfo{person}{Bryan Catanzaro}.}
  \bibinfo{year}{2023}\natexlab{}.
\newblock \showarticletitle{Reducing activation recomputation in large
  transformer models}.
\newblock \bibinfo{journal}{\emph{Proceedings of Machine Learning and Systems}}
   \bibinfo{volume}{5} (\bibinfo{year}{2023}), \bibinfo{pages}{341--353}.
\newblock


\bibitem[Kwon et~al\mbox{.}(2019)]%
        {maestro}
\bibfield{author}{\bibinfo{person}{Hyoukjun Kwon}, \bibinfo{person}{Prasanth
  Chatarasi}, \bibinfo{person}{Michael Pellauer}, \bibinfo{person}{Angshuman
  Parashar}, \bibinfo{person}{Vivek Sarkar}, {and} \bibinfo{person}{Tushar
  Krishna}.} \bibinfo{year}{2019}\natexlab{}.
\newblock \showarticletitle{{Understanding Reuse, Performance, and Hardware
  Cost of DNN Dataflow: A Data-Centric Approach}}. In
  \bibinfo{booktitle}{\emph{Proceedings of the 52nd Annual IEEE/ACM
  International Symposium on Microarchitecture}}. \bibinfo{pages}{754--768}.
\newblock


\bibitem[Kwon et~al\mbox{.}(2023)]%
        {vllm}
\bibfield{author}{\bibinfo{person}{Woosuk Kwon}, \bibinfo{person}{Zhuohan Li},
  \bibinfo{person}{Siyuan Zhuang}, \bibinfo{person}{Ying Sheng},
  \bibinfo{person}{Lianmin Zheng}, \bibinfo{person}{Cody~Hao Yu},
  \bibinfo{person}{Joseph Gonzalez}, \bibinfo{person}{Hao Zhang}, {and}
  \bibinfo{person}{Ion Stoica}.} \bibinfo{year}{2023}\natexlab{}.
\newblock \showarticletitle{{Efficient Memory Management for Large Language
  Model Serving with PagedAttention}}. In \bibinfo{booktitle}{\emph{Proceedings
  of the 29th Symposium on Operating Systems Principles}} (Koblenz, Germany)
  \emph{(\bibinfo{series}{SOSP'23})}.
\newblock
\showISBNx{9798400702297}
\urldef\tempurl%
\url{https://doi.org/10.1145/3600006.3613165}
\showDOI{\tempurl}


\bibitem[Lee et~al\mbox{.}(2021)]%
        {hbm-pim}
\bibfield{author}{\bibinfo{person}{Sukhan Lee}, \bibinfo{person}{Shin-haeng
  Kang}, \bibinfo{person}{Jaehoon Lee}, \bibinfo{person}{Hyeonsu Kim},
  \bibinfo{person}{Eojin Lee}, \bibinfo{person}{Seungwoo Seo},
  \bibinfo{person}{Hosang Yoon}, \bibinfo{person}{Seungwon Lee},
  \bibinfo{person}{Kyounghwan Lim}, \bibinfo{person}{Hyunsung Shin},
  \bibinfo{person}{Jinhyun Kim}, \bibinfo{person}{O Seongil},
  \bibinfo{person}{Anand Iyer}, \bibinfo{person}{David Wang},
  \bibinfo{person}{Kyomin Sohn}, {and} \bibinfo{person}{Nam~Sung Kim}.}
  \bibinfo{year}{2021}\natexlab{}.
\newblock \showarticletitle{{Hardware Architecture and Software Stack for PIM
  Based on Commercial DRAM Technology : Industrial Product}}. In
  \bibinfo{booktitle}{\emph{{2021 ACM/IEEE 48th Annual International Symposium
  on Computer Architecture (ISCA)}}}. \bibinfo{pages}{43--56}.
\newblock


\bibitem[Lee et~al\mbox{.}(2024)]%
        {infinigen}
\bibfield{author}{\bibinfo{person}{Wonbeom Lee}, \bibinfo{person}{Jungi Lee},
  \bibinfo{person}{Junghwan Seo}, {and} \bibinfo{person}{Jaewoong Sim}.}
  \bibinfo{year}{2024}\natexlab{}.
\newblock \showarticletitle{{InfiniGen: Efficient Generative Inference of Large
  Language Models with Dynamic KV Cache Management}}. In
  \bibinfo{booktitle}{\emph{Proceedings of the 18th USENIX Symposium on
  Operating Systems Design and Implementation (OSDI)}}.
\newblock


\bibitem[Li et~al\mbox{.}(2023)]%
        {lightseq}
\bibfield{author}{\bibinfo{person}{Dacheng Li}, \bibinfo{person}{Rulin Shao},
  \bibinfo{person}{Anze Xie}, \bibinfo{person}{Eric~P Xing},
  \bibinfo{person}{Joseph~E Gonzalez}, \bibinfo{person}{Ion Stoica},
  \bibinfo{person}{Xuezhe Ma}, {and} \bibinfo{person}{Hao Zhang}.}
  \bibinfo{year}{2023}\natexlab{}.
\newblock \showarticletitle{Lightseq: Sequence level parallelism for
  distributed training of long context transformers}.
\newblock \bibinfo{journal}{\emph{arXiv preprint arXiv:2310.03294}}
  (\bibinfo{year}{2023}).
\newblock


\bibitem[Li et~al\mbox{.}(2021)]%
        {seqparallelism1}
\bibfield{author}{\bibinfo{person}{Shenggui Li}, \bibinfo{person}{Fuzhao Xue},
  \bibinfo{person}{Chaitanya Baranwal}, \bibinfo{person}{Yongbin Li}, {and}
  \bibinfo{person}{Yang You}.} \bibinfo{year}{2021}\natexlab{}.
\newblock \showarticletitle{Sequence parallelism: Long sequence training from
  system perspective}.
\newblock \bibinfo{journal}{\emph{arXiv preprint arXiv:2105.13120}}
  (\bibinfo{year}{2021}).
\newblock


\bibitem[Li et~al\mbox{.}(2020)]%
        {dramsim3}
\bibfield{author}{\bibinfo{person}{S. Li}, \bibinfo{person}{Z. Yang},
  \bibinfo{person}{D. Reddy}, \bibinfo{person}{A. Srivastava}, {and}
  \bibinfo{person}{B. Jacob}.} \bibinfo{year}{2020}\natexlab{}.
\newblock \showarticletitle{{DRAMsim3: A Cycle-Accurate, Thermal-Capable DRAM
  Simulator}}.
\newblock  \bibinfo{volume}{19}, \bibinfo{number}{2} (\bibinfo{year}{2020}),
  \bibinfo{pages}{106--109}.
\newblock


\bibitem[NVIDIA(2006)]%
        {cuda-event}
\bibfield{author}{\bibinfo{person}{NVIDIA}.} \bibinfo{year}{2006}\natexlab{}.
\newblock \bibinfo{title}{{CUDA Toolkit Document}}.
\newblock
  \bibinfo{howpublished}{\url{https://docs.nvidia.com/cuda/cuda-runtime-api}}.
\newblock


\bibitem[NVIDIA(2014)]%
        {cuda-um}
\bibfield{author}{\bibinfo{person}{NVIDIA}.} \bibinfo{year}{2014}\natexlab{}.
\newblock \bibinfo{title}{{Unified Memory Programming}}.
\newblock
  \bibinfo{howpublished}{https://docs.nvidia.com/cuda/cuda-c-programming-guide/index.html\#um-unified-memory-programminghd}.
\newblock


\bibitem[NVIDIA(2023)]%
        {gracehopper}
\bibfield{author}{\bibinfo{person}{NVIDIA}.} \bibinfo{year}{2023}\natexlab{}.
\newblock \bibinfo{title}{{Grace Hopper Superchip}}.
\newblock
  \bibinfo{howpublished}{https://www.nvidia.com/en-us/data-center/grace-hopper-superchip/}.
\newblock


\bibitem[NVIDIA(2024)]%
        {blackwell}
\bibfield{author}{\bibinfo{person}{NVIDIA}.} \bibinfo{year}{2024}\natexlab{}.
\newblock \bibinfo{title}{{Blackwell Superchip}}.
\newblock
  \bibinfo{howpublished}{https://www.nvidia.com/en-us/data-center/gb200-nvl72/}.
\newblock


\bibitem[OpenAI(2023)]%
        {gpt4}
\bibfield{author}{\bibinfo{person}{OpenAI}.} \bibinfo{year}{2023}\natexlab{}.
\newblock \showarticletitle{{GPT-4 Technical Report}}.
\newblock \bibinfo{journal}{\emph{arXiv preprint arXiv:2303. 08774}}
  (\bibinfo{year}{2023}).
\newblock
\showeprint[arxiv]{2303.08774}~[cs.CL]


\bibitem[Park et~al\mbox{.}(2024a)]%
        {attacc}
\bibfield{author}{\bibinfo{person}{Jaehyun Park}, \bibinfo{person}{Jaewan
  Choi}, \bibinfo{person}{Kwanhee Kyung}, \bibinfo{person}{Michael~Jaemin Kim},
  \bibinfo{person}{Yongsuk Kwon}, \bibinfo{person}{Nam~Sung Kim}, {and}
  \bibinfo{person}{Jung~Ho Ahn}.} \bibinfo{year}{2024}\natexlab{a}.
\newblock \showarticletitle{{AttAcc! Unleashing the Power of PIM for Batched
  Transformer-based Generative Model Inference}}. In
  \bibinfo{booktitle}{\emph{Proceedings of the 29th ACM International
  Conference on Architectural Support for Programming Languages and Operating
  Systems (ASPLOS'24)}}.
\newblock


\bibitem[Park et~al\mbox{.}(2024b)]%
        {cxlpnm}
\bibfield{author}{\bibinfo{person}{Sang-Soo Park}, \bibinfo{person}{KyungSoo
  Kim}, \bibinfo{person}{Jinin So}, \bibinfo{person}{Jin Jung},
  \bibinfo{person}{Jonggeon Lee}, \bibinfo{person}{Kyoungwan Woo},
  \bibinfo{person}{Nayeon Kim}, \bibinfo{person}{Younghyun Lee},
  \bibinfo{person}{Hyungyo Kim}, \bibinfo{person}{Yongsuk Kwon},
  {et~al\mbox{.}}} \bibinfo{year}{2024}\natexlab{b}.
\newblock \showarticletitle{An LPDDR-based CXL-PNM Platform for TCO-efficient
  Inference of Transformer-based Large Language Models}. In
  \bibinfo{booktitle}{\emph{2024 IEEE International Symposium on
  High-Performance Computer Architecture (HPCA)}}. IEEE,
  \bibinfo{pages}{970--982}.
\newblock


\bibitem[Pope et~al\mbox{.}(2023)]%
        {kvcache}
\bibfield{author}{\bibinfo{person}{Reiner Pope}, \bibinfo{person}{Sholto
  Douglas}, \bibinfo{person}{Aakanksha Chowdhery}, \bibinfo{person}{Jacob
  Devlin}, \bibinfo{person}{James Bradbury}, \bibinfo{person}{Jonathan Heek},
  \bibinfo{person}{Kefan Xiao}, \bibinfo{person}{Shivani Agrawal}, {and}
  \bibinfo{person}{Jeff Dean}.} \bibinfo{year}{2023}\natexlab{}.
\newblock \showarticletitle{Efficiently scaling transformer inference}.
\newblock \bibinfo{journal}{\emph{Proceedings of Machine Learning and Systems}}
   \bibinfo{volume}{5} (\bibinfo{year}{2023}).
\newblock


\bibitem[Prabhu et~al\mbox{.}(2024)]%
        {vattention}
\bibfield{author}{\bibinfo{person}{Ramya Prabhu}, \bibinfo{person}{Ajay Nayak},
  \bibinfo{person}{Jayashree Mohan}, \bibinfo{person}{Ramachandran Ramjee},
  {and} \bibinfo{person}{Ashish Panwar}.} \bibinfo{year}{2024}\natexlab{}.
\newblock \showarticletitle{vAttention: Dynamic Memory Management for Serving
  LLMs without PagedAttention}.
\newblock \bibinfo{journal}{\emph{arXiv preprint arXiv:2405.04437}}
  (\bibinfo{year}{2024}).
\newblock


\bibitem[Rhu et~al\mbox{.}(2016)]%
        {vdnn}
\bibfield{author}{\bibinfo{person}{Minsoo Rhu}, \bibinfo{person}{Natalia
  Gimelshein}, \bibinfo{person}{Jason Clemons}, \bibinfo{person}{Arslan
  Zulfiqar}, {and} \bibinfo{person}{Stephen~W Keckler}.}
  \bibinfo{year}{2016}\natexlab{}.
\newblock \showarticletitle{vDNN: Virtualized Deep Neural Networks for
  Scalable, Memory-Efficient Neural Network Design}. In
  \bibinfo{booktitle}{\emph{2016 49th Annual IEEE/ACM International Symposium
  on Microarchitecture (MICRO)}}. IEEE, \bibinfo{pages}{1--13}.
\newblock


\bibitem[Scao et~al\mbox{.}(2023)]%
        {bloom}
\bibfield{author}{\bibinfo{person}{Teven~Le Scao}, \bibinfo{person}{Angela
  Fan}, \bibinfo{person}{Christopher Akiki}, \bibinfo{person}{Ellie Pavlick},
  \bibinfo{person}{Suzana Ilić}, \bibinfo{person}{Daniel Hesslow},
  \bibinfo{person}{Roman Castagné}, \bibinfo{person}{Alexandra~Sasha
  Luccioni}, \bibinfo{person}{François Yvon}, \bibinfo{person}{Matthias
  Gallé}, {et~al\mbox{.}}} \bibinfo{year}{2023}\natexlab{}.
\newblock \showarticletitle{{BLOOM: A 176B-Parameter Open-Access Multilingual
  Language Model}}.
\newblock \bibinfo{journal}{\emph{arXiv preprint arXiv:2211.05100}}
  (\bibinfo{year}{2023}).
\newblock
\showeprint[arxiv]{2211.05100}~[cs.CL]


\bibitem[Seo et~al\mbox{.}(2024)]%
        {ianus}
\bibfield{author}{\bibinfo{person}{Minseok Seo}, \bibinfo{person}{Xuan~Truong
  Nguyen}, \bibinfo{person}{Seok~Joong Hwang}, \bibinfo{person}{Yongkee Kwon},
  \bibinfo{person}{Guhyun Kim}, \bibinfo{person}{Chanwook Park},
  \bibinfo{person}{Ilkon Kim}, \bibinfo{person}{Jaehan Park},
  \bibinfo{person}{Jeongbin Kim}, \bibinfo{person}{Woojae Shin},
  \bibinfo{person}{Jongsoon Won}, \bibinfo{person}{Haerang Choi},
  \bibinfo{person}{Kyuyoung Kim}, \bibinfo{person}{Daehan Kwon},
  \bibinfo{person}{Chunseok Jeong}, \bibinfo{person}{Sangheon Lee},
  \bibinfo{person}{Yongseok Choi}, \bibinfo{person}{Wooseok Byun},
  \bibinfo{person}{Seungcheol Baek}, \bibinfo{person}{Hyuk-Jae Lee}, {and}
  \bibinfo{person}{John Kim}.} \bibinfo{year}{2024}\natexlab{}.
\newblock \showarticletitle{{IANUS: Integrated Accelerator based on NPU-PIM
  Unified Memory System}}. In \bibinfo{booktitle}{\emph{Proceedings of the 29th
  ACM International Conference on Architectural Support for Programming
  Languages and Operating Systems (ASPLOS'24)}}.
\newblock


\bibitem[Sheng et~al\mbox{.}(2023)]%
        {flexgen}
\bibfield{author}{\bibinfo{person}{Ying Sheng}, \bibinfo{person}{Lianmin
  Zheng}, \bibinfo{person}{Binhang Yuan}, \bibinfo{person}{Zhuohan Li},
  \bibinfo{person}{Max Ryabinin}, \bibinfo{person}{Beidi Chen},
  \bibinfo{person}{Percy Liang}, \bibinfo{person}{Christopher R\'{e}},
  \bibinfo{person}{Ion Stoica}, {and} \bibinfo{person}{Ce Zhang}.}
  \bibinfo{year}{2023}\natexlab{}.
\newblock \showarticletitle{{FlexGen: High-Throughput Generative Inference of
  Large Language Models with a Single GPU}}. In
  \bibinfo{booktitle}{\emph{Proceedings of the 40th International Conference on
  Machine Learning}} (Honolulu, Hawaii, USA)
  \emph{(\bibinfo{series}{ICML'23})}.
\newblock


\bibitem[Shoeybi et~al\mbox{.}(2019)]%
        {megatron}
\bibfield{author}{\bibinfo{person}{Mohammad Shoeybi}, \bibinfo{person}{Mostofa
  Patwary}, \bibinfo{person}{Raul Puri}, \bibinfo{person}{Patrick LeGresley},
  \bibinfo{person}{Jared Casper}, {and} \bibinfo{person}{Bryan Catanzaro}.}
  \bibinfo{year}{2019}\natexlab{}.
\newblock \showarticletitle{Megatron-lm: Training multi-billion parameter
  language models using model parallelism}.
\newblock \bibinfo{journal}{\emph{arXiv preprint arXiv:1909.08053}}
  (\bibinfo{year}{2019}).
\newblock


\bibitem[Song et~al\mbox{.}(2024)]%
        {powerinfer}
\bibfield{author}{\bibinfo{person}{Yixin Song}, \bibinfo{person}{Zeyu Mi},
  \bibinfo{person}{Haotong Xie}, {and} \bibinfo{person}{Haibo Chen}.}
  \bibinfo{year}{2024}\natexlab{}.
\newblock \showarticletitle{{PowerInfer: Fast Large Language Model Serving with
  a Consumer-grade GPU}}. In \bibinfo{booktitle}{\emph{Proceedings of the 30th
  Symposium on Operating Systems Principles (SOSP)}}.
\newblock


\bibitem[Touvron et~al\mbox{.}(2023a)]%
        {llama}
\bibfield{author}{\bibinfo{person}{Hugo Touvron}, \bibinfo{person}{Thibaut
  Lavril}, \bibinfo{person}{Gautier Izacard}, \bibinfo{person}{Xavier
  Martinet}, \bibinfo{person}{Marie-Anne Lachaux},
  \bibinfo{person}{Timoth{\'e}e Lacroix}, \bibinfo{person}{Baptiste
  Rozi{\`e}re}, \bibinfo{person}{Naman Goyal}, \bibinfo{person}{Eric Hambro},
  \bibinfo{person}{Faisal Azhar}, {et~al\mbox{.}}}
  \bibinfo{year}{2023}\natexlab{a}.
\newblock \showarticletitle{{Llama: Open and efficient foundation language
  models}}.
\newblock \bibinfo{journal}{\emph{arXiv preprint arXiv:2302.13971}}
  (\bibinfo{year}{2023}).
\newblock


\bibitem[Touvron et~al\mbox{.}(2023b)]%
        {llama2}
\bibfield{author}{\bibinfo{person}{Hugo Touvron}, \bibinfo{person}{Louis
  Martin}, \bibinfo{person}{Kevin Stone}, \bibinfo{person}{Peter Albert},
  \bibinfo{person}{Amjad Almahairi}, \bibinfo{person}{Yasmine Babaei},
  \bibinfo{person}{Nikolay Bashlykov}, \bibinfo{person}{Soumya Batra},
  \bibinfo{person}{Prajjwal Bhargava}, \bibinfo{person}{Shruti Bhosale},
  \bibinfo{person}{Dan Bikel}, \bibinfo{person}{Lukas Blecher},
  \bibinfo{person}{Cristian~Canton Ferrer}, \bibinfo{person}{Moya Chen},
  \bibinfo{person}{Guillem Cucurull}, \bibinfo{person}{David Esiobu},
  \bibinfo{person}{Jude Fernandes}, \bibinfo{person}{Jeremy Fu},
  \bibinfo{person}{Wenyin Fu}, \bibinfo{person}{Brian Fuller},
  \bibinfo{person}{Cynthia Gao}, \bibinfo{person}{Vedanuj Goswami},
  \bibinfo{person}{Naman Goyal}, \bibinfo{person}{Anthony Hartshorn},
  \bibinfo{person}{Saghar Hosseini}, \bibinfo{person}{Rui Hou},
  \bibinfo{person}{Hakan Inan}, \bibinfo{person}{Marcin Kardas},
  \bibinfo{person}{Viktor Kerkez}, \bibinfo{person}{Madian Khabsa},
  \bibinfo{person}{Isabel Kloumann}, \bibinfo{person}{Artem Korenev},
  \bibinfo{person}{Punit~Singh Koura}, \bibinfo{person}{Marie-Anne Lachaux},
  \bibinfo{person}{Thibaut Lavril}, \bibinfo{person}{Jenya Lee},
  \bibinfo{person}{Diana Liskovich}, \bibinfo{person}{Yinghai Lu},
  \bibinfo{person}{Yuning Mao}, \bibinfo{person}{Xavier Martinet},
  \bibinfo{person}{Todor Mihaylov}, \bibinfo{person}{Pushkar Mishra},
  \bibinfo{person}{Igor Molybog}, \bibinfo{person}{Yixin Nie},
  \bibinfo{person}{Andrew Poulton}, \bibinfo{person}{Jeremy Reizenstein},
  \bibinfo{person}{Rashi Rungta}, \bibinfo{person}{Kalyan Saladi},
  \bibinfo{person}{Alan Schelten}, \bibinfo{person}{Ruan Silva},
  \bibinfo{person}{Eric~Michael Smith}, \bibinfo{person}{Ranjan Subramanian},
  \bibinfo{person}{Xiaoqing~Ellen Tan}, \bibinfo{person}{Binh Tang},
  \bibinfo{person}{Ross Taylor}, \bibinfo{person}{Adina Williams},
  \bibinfo{person}{Jian~Xiang Kuan}, \bibinfo{person}{Puxin Xu},
  \bibinfo{person}{Zheng Yan}, \bibinfo{person}{Iliyan Zarov},
  \bibinfo{person}{Yuchen Zhang}, \bibinfo{person}{Angela Fan},
  \bibinfo{person}{Melanie Kambadur}, \bibinfo{person}{Sharan Narang},
  \bibinfo{person}{Aurelien Rodriguez}, \bibinfo{person}{Robert Stojnic},
  \bibinfo{person}{Sergey Edunov}, {and} \bibinfo{person}{Thomas Scialom}.}
  \bibinfo{year}{2023}\natexlab{b}.
\newblock \bibinfo{title}{{Llama 2: Open Foundation and Fine-Tuned Chat
  Models}}.
\newblock
\newblock
\showeprint[arxiv]{2307.09288}~[cs.CL]


\bibitem[Vaswani et~al\mbox{.}(2017)]%
        {attention}
\bibfield{author}{\bibinfo{person}{Ashish Vaswani}, \bibinfo{person}{Noam
  Shazeer}, \bibinfo{person}{Niki Parmar}, \bibinfo{person}{Jakob Uszkoreit},
  \bibinfo{person}{Llion Jones}, \bibinfo{person}{Aidan~N. Gomez},
  \bibinfo{person}{Lukasz Kaiser}, {and} \bibinfo{person}{Illia Polosukhin}.}
  \bibinfo{year}{2017}\natexlab{}.
\newblock \showarticletitle{{Attention Is All You Need}}. In
  \bibinfo{booktitle}{\emph{31st Conference on Neural Information Processing
  Systems (NIPS 2017)}}.
\newblock


\bibitem[Wu et~al\mbox{.}(2019)]%
        {accelergy}
\bibfield{author}{\bibinfo{person}{Yannan~Nellie Wu}, \bibinfo{person}{Joel~S
  Emer}, {and} \bibinfo{person}{Vivienne Sze}.}
  \bibinfo{year}{2019}\natexlab{}.
\newblock \showarticletitle{Accelergy: An architecture-level energy estimation
  methodology for accelerator designs}. In \bibinfo{booktitle}{\emph{2019
  IEEE/ACM International Conference on Computer-Aided Design (ICCAD)}}.
  \bibinfo{pages}{1--8}.
\newblock


\bibitem[Yu et~al\mbox{.}(2022)]%
        {orca}
\bibfield{author}{\bibinfo{person}{Gyeong-In Yu}, \bibinfo{person}{Joo~Seong
  Jeong}, \bibinfo{person}{Geon-Woo Kim}, \bibinfo{person}{Soojeong Kim}, {and}
  \bibinfo{person}{Byung-Gon Chun}.} \bibinfo{year}{2022}\natexlab{}.
\newblock \showarticletitle{{Orca: A distributed serving system for
  Transformer-Based generative models}}. In \bibinfo{booktitle}{\emph{16th
  USENIX Symposium on Operating Systems Design and Implementation (OSDI 22)}}.
  \bibinfo{pages}{521--538}.
\newblock


\bibitem[Zeng et~al\mbox{.}(2024)]%
        {flightllm}
\bibfield{author}{\bibinfo{person}{Shulin Zeng}, \bibinfo{person}{Jun Liu},
  \bibinfo{person}{Guohao Dai}, \bibinfo{person}{Xinhao Yang},
  \bibinfo{person}{Tianyu Fu}, \bibinfo{person}{Hongyi Wang},
  \bibinfo{person}{Wenheng Ma}, \bibinfo{person}{Hanbo Sun},
  \bibinfo{person}{Shiyao Li}, \bibinfo{person}{Zixiao Huang}, {et~al\mbox{.}}}
  \bibinfo{year}{2024}\natexlab{}.
\newblock \showarticletitle{FlightLLM: Efficient Large Language Model Inference
  with a Complete Mapping Flow on FPGA}.
\newblock \bibinfo{journal}{\emph{arXiv preprint arXiv:2401.03868}}
  (\bibinfo{year}{2024}).
\newblock


\bibitem[Zhang et~al\mbox{.}(2022)]%
        {opt}
\bibfield{author}{\bibinfo{person}{Susan Zhang}, \bibinfo{person}{Stephen
  Roller}, \bibinfo{person}{Naman Goyal}, \bibinfo{person}{Mikel Artetxe},
  \bibinfo{person}{Moya Chen}, \bibinfo{person}{Shuohui Chen},
  \bibinfo{person}{Christopher Dewan}, \bibinfo{person}{Mona Diab},
  \bibinfo{person}{Xian Li}, \bibinfo{person}{Xi~Victoria Lin},
  \bibinfo{person}{Todor Mihaylov}, \bibinfo{person}{Myle Ott},
  \bibinfo{person}{Sam Shleifer}, \bibinfo{person}{Kurt Shuster},
  \bibinfo{person}{Daniel Simig}, \bibinfo{person}{Punit~Singh Koura},
  \bibinfo{person}{Anjali Sridhar}, \bibinfo{person}{Tianlu Wang}, {and}
  \bibinfo{person}{Luke Zettlemoyer}.} \bibinfo{year}{2022}\natexlab{}.
\newblock \bibinfo{title}{{OPT: Open Pre-trained Transformer Language Models}}.
\newblock
\newblock
\showeprint[arxiv]{2205.01068}~[cs.CL]


\bibitem[Zhao et~al\mbox{.}(2023)]%
        {llm-survey}
\bibfield{author}{\bibinfo{person}{Wayne~Xin Zhao}, \bibinfo{person}{Kun Zhou},
  \bibinfo{person}{Junyi Li}, \bibinfo{person}{Tianyi Tang},
  \bibinfo{person}{Xiaolei Wang}, \bibinfo{person}{Yupeng Hou},
  \bibinfo{person}{Yingqian Min}, \bibinfo{person}{Beichen Zhang},
  \bibinfo{person}{Junjie Zhang}, \bibinfo{person}{Zican Dong},
  \bibinfo{person}{Yifan Du}, \bibinfo{person}{Chen Yang},
  \bibinfo{person}{Yushuo Chen}, \bibinfo{person}{Zhipeng Chen},
  \bibinfo{person}{Jinhao Jiang}, \bibinfo{person}{Ruiyang Ren},
  \bibinfo{person}{Yifan Li}, \bibinfo{person}{Xinyu Tang},
  \bibinfo{person}{Zikang Liu}, \bibinfo{person}{Peiyu Liu},
  \bibinfo{person}{Jian-Yun Nie}, {and} \bibinfo{person}{Ji-Rong Wen}.}
  \bibinfo{year}{2023}\natexlab{}.
\newblock \showarticletitle{{A Survey on Large Language Models}}.
\newblock \bibinfo{journal}{\emph{arXiv preprint arXiv:2303.18223}}
  (\bibinfo{year}{2023}).
\newblock
\showeprint[arxiv]{2303.18223}~[cs.CL]


\bibitem[Zhao et~al\mbox{.}(2024)]%
        {alisa}
\bibfield{author}{\bibinfo{person}{Youpeng Zhao}, \bibinfo{person}{Di Wu},
  {and} \bibinfo{person}{Jun Wang}.} \bibinfo{year}{2024}\natexlab{}.
\newblock \showarticletitle{ALISA: Accelerating Large Language Model Inference
  via Sparsity-Aware KV Caching}. In \bibinfo{booktitle}{\emph{2024 ACM/IEEE
  51st Annual International Symposium on Computer Architecture (ISCA)}}.
\newblock


\bibitem[Zheng et~al\mbox{.}(2022)]%
        {alpa}
\bibfield{author}{\bibinfo{person}{Lianmin Zheng}, \bibinfo{person}{Zhuohan
  Li}, \bibinfo{person}{Hao Zhang}, \bibinfo{person}{Yonghao Zhuang},
  \bibinfo{person}{Zhifeng Chen}, \bibinfo{person}{Yanping Huang},
  \bibinfo{person}{Yida Wang}, \bibinfo{person}{Yuanzhong Xu},
  \bibinfo{person}{Danyang Zhuo}, \bibinfo{person}{Eric~P. Xing},
  \bibinfo{person}{Joseph~E. Gonzalez}, {and} \bibinfo{person}{Ion Stoica}.}
  \bibinfo{year}{2022}\natexlab{}.
\newblock \showarticletitle{{Alpa: Automating Inter- and {Intra-Operator}
  Parallelism for Distributed Deep Learning}}. In
  \bibinfo{booktitle}{\emph{16th USENIX Symposium on Operating Systems Design
  and Implementation (OSDI 22)}}. \bibinfo{publisher}{USENIX Association},
  \bibinfo{address}{Carlsbad, CA}.
\newblock
\showISBNx{978-1-939133-28-1}
\urldef\tempurl%
\url{https://www.usenix.org/conference/osdi22/presentation/zheng-lianmin}
\showURL{%
\tempurl}


\end{thebibliography}

\end{document}